 \documentclass[manuscript]{emulateapj-rtx4}

\newcommand{\rhoGJ}{\rho_{{\rm GJ}}}  
\newcommand{\Ell}{E_\parallel}      
\newcommand{\rhowSQR}{\rho_{\rm w}^2}

\slugcomment{}

\shorttitle{Lepton acceleration in the vicinity of the event horizon}
\shortauthors{Hirotani et al.}

\begin{document}


\title{Lepton acceleration in the vicinity of the event horizon:
       High-energy and Very-high-energy emissions from
       rotating black holes with various masses}


 \author{Kouichi Hirotani${}^1$, 
         Hung-Yi Pu${}^1$, 
         Lupin Chun-Che Lin${}^1$,
         Hsiang-Kuang Chang${}^2$
         Makoto Inoue${}^1$,
         Albert K. H Kong${}^2$,
         Satoki Matsushita${}^1$, and
         Pak-Hin T. Tam${}^3$
         }
 \affil{${}^1$
        Academia Sinica, Institute of Astronomy and Astrophysics (ASIAA),
        PO Box 23-141, Taipei, Taiwan 10617, R.O.C.;
        hirotani@tiara.sinica.edu.tw}
\affil{${}^2$
       Institute of Astronomy,
       Department of Physics, 
       National Tsing Hua University,
       No. 101, Section 2, Kuang-Fu Road, Hsinchu, Taiwan 30013, R.O.C.}
 \affil{${}^3$
       School of Physics and Astronomy, Sun Yat-Sen University, 
       Zhuhai 519082, China}

%
%
%
%


\begin{abstract}
We investigate the electrostatic acceleration of electrons and positrons 
in the vicinity of the event horizon,
applying the pulsar outer-gap model to black hole magnetospheres. 
During a low accretion phase,
the radiatively inefficient accretion flow (RIAF) cannot emit
enough MeV photons that are needed to sustain the force-free 
magnetosphere via two-photon collisions.
In such a charge-starved region (or a gap), an electric field arises 
along the magnetic field lines to accelerate electrons 
and positrons into ultra-relativistic energies.
These relativistic leptons 
emit copious gamma-rays via curvature and inverse-Compton (IC) processes.
Some of such gamma-rays collide with the submillimeter-IR photons 
emitted from the RIAF to materialize as pairs,
which polarize to partially screen the original acceleration 
electric field.
It is found that the gap gamma-ray luminosity 
increases with decreasing accretion rate.
However, if the accretion rate decreases too much, 
the diminished RIAF soft photon field can no longer sustain
a stationary pair production within the gap.
As long as a stationary gap is formed, 
the magnetosphere becomes force-free outside the gap
by the cascaded pairs, irrespective of the BH mass.
If a nearby stellar-mass black hole (BH) is in quiescence,
or if a galactic intermediate-mass BH is in a very low accretion state,
its curvature and IC emissions are 
found to be detectable with {\it Fermi}/LAT 
and imaging atmospheric Cherenkov telescopes (IACT).
If a low-luminosity active galactic nucleus is 
located within a few tens of Mpc, 
the IC emission from its super-massive BH
is marginally detectable with IACT.

\end{abstract}





\section{Introduction}
%
%


It is widely accepted that an accreting black hole (BH) 
of an arbitrary size can produce a pair of relativistic plasma jets, 
which are often observed in various wavelengths
from radio to very-high-energy (VHE) $\gamma$-rays.
The most plausible mechanism for powering such jets is the extraction 
of the BH rotational energy through the Blandford-Znajek (BZ) process
\citep{bla76}.
In this electromagnetic process, the magnetic field lines
threading the event horizon exert a counter torque on it
to spin down the BH,
launching Poynting-flux-dominated outflows \citep{koide02}.
Indeed, general relativistic (GR) magnetohydrodynamic (MHD) models 
show the existence of 
nearly steady, collimated, magnetically dominated jets
in the polar regions
\citep{mckinney04,hirose04},
whose structures are similar to those 
in the force-free models
\citep{hawley06,mckinney07a,mckinney07b}.
Since the centrifugal-force barrier prevents plasma accretion 
towards the rotation axis,
the magnetic energy density dominates the plasmas' 
rest-mass energy density in these polar funnels. 

Even if the axial funnels are almost force-free 
in this sense,
it is the electric current that sustains the electromagnetic power, 
and it is the charged particles that carry the electric current.
That is, even under the assumption of masslessness, 
it must be the real charged particles that constitute 
the electric currents.
In the direct vicinity of the horizon,
causality requires that 
positive (or negative) charges must flow into the horizon 
when an electric current flows into (or out from) the horizon.
Since accreting plasmas cannot easily penetrate into the funnels,
and since they cannot emit sufficient MeV photons
that are capable of materializing as electron-positron pairs
when the accretion rate is very low,
we need a process of plasma supply deep within the
magnetosphere somewhere above the horizon.

To contrive a plasma source in the vicinity of the horizon,
\citet{bes92} extended the pulsar outer-magnetospheric 
lepton accelerator model \citep{cheng86a,cheng86b} 
to BH magnetospheres, and proposed the BH gap model.
Extending this pioneering work, \citet{hiro98} demonstrated that
a sufficient electric current can be supplied 
via copious pair production around super-massive BHs (SMBHs).
However, the predicted $\gamma$-ray luminosity was too small
to be detected by any instruments,
because they assumed a substantial accretion rate (as in quasars),
which leads to a very thin gap width along the magnetic field lines.

On these grounds, \citet{nero07} and \citet{levi11} 
revisited the BH gap model
and applied it the central BH of radio galaxy M87 (i.e., M87*) 
and Srg~A*,
adopting a much thicker gap width as large as the horizon radius.
Their assumption of a thick gap was, indeed, reasonable,
because such a low luminosity active galactic nucleus
generally possesses a less dense radiation field in the center
and hence a geometrically extended gap slightly above the horizon.
Then \citet{brod15} demonstrated that the two-stream instability 
does not grow in the ultra-relativistic, counter-streaming pairs 
in BH gaps.
Subsequently, \citet[][hereafter HP16]{hiro16}
showed that a gap arises around 
the so-called \lq null-charge surface',
on which the GR Goldreich-Julian (GJ) charge density vanishes,
by solving the set of 
an inhomogeneous part of the Maxwell equations,
lepton equations of motion, and 
the radiative transfer equation.
Then they applied their BH-gap model to the radio galaxy IC310,
whose central BH (i.e., IC310*) is accreting plasmas 
at much smaller rate than the Eddington rate.
They demonstrated 
that the gap width becomes as large as the horizon radius
when the accretion rate is very low,
and that the observed VHE flux during the flare
cannot be reproduced by their BH gap model,
unless the magnetic field becomes much stronger than
the equipartition value with the plasma accretions.
Noting that GR effects most importantly appear in the formation
of the null-charge surface through frame dragging,
they evaluated the GJ charge density general relativistically.
However, they neglected the GR effects in any other terms of
the basic equations as the first step,
because the electromagnetic interaction dominates the gravitational
one except for the direct vicinity of the horizon,
and because the photons emitted from the horizon vicinity
will not strongly affect the emission spectra due to redshift.

In the present paper, 
to further quantify the gap model,
we include the GR effects in all the basic equations and calculations,
extending the method of HP16.
In the next section,
we examine the detectability of BH gap emissions
for various BH masses. 
Then in \S~\ref{sec:geometry},
we describe the background space time and derive the
GR GJ charge density in a rotating BH magnetosphere.
We then formulate the basic equations of gap electrodynamics
in \S~\ref{sec:accelerator},
and investigate their emission properties in \S~\ref{sec:solutions}.
In the final section, 
we discuss how to discriminate the gap emission from the jet emission.

\section{Detectability of gap emissions}
\label{sec:detect}
Since the gap liberate only a portion of the electromagnetic power
extracted from a rotating BH,
the upper limit of its luminosity can be given by the
Blandford-Znajek power \citep{bla76}
\begin{equation}
  L_{\rm BZ} 
    \approx 10^{21} a_\ast{}^2 M_1{}^2 B^2
    \mbox{erg s}^{-1},
  \label{eq:L_BZ}
\end{equation}
where $a_\ast \equiv a/r_{\rm g}$ denotes the dimensionless BH's 
spin parameter,
$a \equiv J/(Mc)$ the spin parameter,
$J$ the BH's angular momentum,
$M$ the BH mass, 
$c$ the speed of light,
$r_{\rm g} \equiv GM c^{-2}$ the gravitational radius, 
$G$ the gravitational constant, 
$B$ the magnetic field strength in gauss, and
$M_1 \equiv M/(10 M_\odot)$.

In the present paper, unless explicitly mentioned,
we assume that the magnetic buoyancy balances disk gravity
and evaluate $B$ with 
the equi-partition value.
Near the horizon, we obtain
\citep{bisno74,gosh77,narayan03,levi11}
\begin{equation}
  B_{\rm eq} 
    \approx 4 \times 10^8 \dot{m}^{1/2} M_1{}^{-1/2} 
            \mbox{ G},
  \label{eq:B_eq}
\end{equation}
where $\dot{m}$ refers to the dimensionless accretion rate
near the horizon and is defined as
$ \dot{m} \equiv \dot{M} / \dot{M}_{\rm Edd}$;
$\dot{M}$ denotes the mass accretion rate, 
$\dot{M}_{\rm Edd} \equiv L_{\rm Edd} / \eta c^2$,
$L_{\rm Edd}=1.25 \times 10^{39} M_1 \mbox{ ergs s}^{-1}$,
and $\eta \approx 0.1$ \citep{mckinney12}.
Substituting $B=B_{\rm eq}$ into equation~(\ref{eq:L_BZ}), 
we obtain 
\begin{equation}
  L_{\rm BZ} 
    \approx 1.7 \times 10^{38} a_\ast{}^2 
                       \dot{m} M_1 
    \mbox{erg s}^{-1}.
  \label{eq:L_BZ_2}
\end{equation}

\begin{figure}
 \includegraphics[angle=0,scale=0.40]{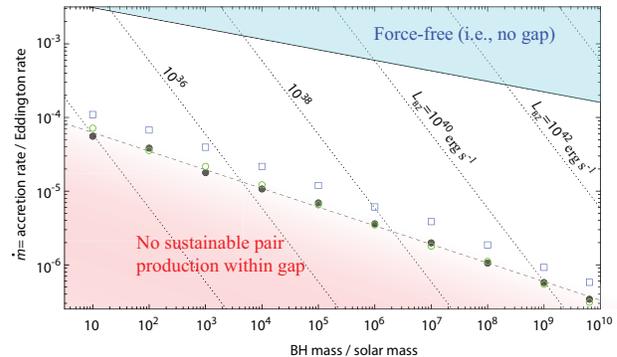}
\caption{
Dimensionless accretion rate, $\dot{m}$, versus black hole mass, $M$.
The thick solid line shows the upper limit of $\dot{m}$
above which the copious pair production by the RIAF emission
prevents the formation of a gap 
(and ensures the force-free magnetosphere).
The filled circles, open circles, 
and open squares denote the solved
lower limits of $\dot{m}$ (\S~\ref{sec:general}) for 
$(a_\ast,\Omega_{\rm F}/\omega_{\rm H},R_{\rm min}/r_{\rm g})
 =(0.9,0.5,6)$,
$(0.5,0.5,6)$, and
$(0.9,0.5,12)$, respectively.
The thin dashed line shows a linear fit of the filled circles,
$\lg \dot{m}_{\rm low}= -3.64 -0.261 \lg (M/M_\odot)$.
The dotted lines show the extracted power from the black hole
(eq.~\ref{eq:L_BZ}) for an extremely rotating case, $a=M$.
For the explanation of the red-graded region, 
see the end of \S~\ref{sec:SMBH}.
\label{fig:mdot}
}
\end{figure}

The magnetosphere becomes force-free,
that is, an electric field does not arise along the magnetic field lines,
if the pair density exceeds the GJ value,
$N_{\rm GJ} \sim \Omega_{\rm F} B/(2\pi ce)$, 
where $\Omega_{\rm F}$ denotes the angular frequency 
of the magnetic field and $e$ the magnitude of the
charge on the electron.
In the vicinity of an accreting BH,
a plasma accretion becomes radiatively inefficient
when the accretion rate typically reduces to $\dot{m} < 10^{-2.5}$
\citep{ichimaru77,narayan94,narayan95}.
From such a radiatively inefficient accretion flow (RIAF),
MeV photons are emitted via free-free process,
colliding each other to materialize as electron-positron pairs
in the magnetosphere.
If the number density, $N_\pm$, of such created pairs 
becomes less than $N_{\rm GJ}$, 
charges cannot completely screen an electric field, $E_\parallel$, 
along the magnetic field line;  
that is, a gap (i.e., a charge-starved region) appears.
The RIAF theory gives $N_\pm$ as a function of
$M$ and $\dot{m}$ \citep{levi11}.
Thus, putting $N_\pm < N_{\rm GJ}$,
we obtain a condition for a gap to appear,
\begin{equation}
  \dot{m} 
    < \dot{m}_{\rm up} \equiv 3.1 \times 10^{-3} M_1{}^{-1/7} .
  \label{eq:mdot_up}
\end{equation}
In figure~\ref{fig:mdot}, we plot this upper limit,
$\dot{m}_{\rm up}=\dot{m}_{\rm up}(M)$
as the thick solid line.
The dotted lines denote $L_{\rm BZ}$ (eq.~[\ref{eq:L_BZ_2}])
as labeled. 
The crossing of the solid line and a dotted line gives
the Blandford-Znajek power at each BH mass.
For example, the thick solid line and 
the dotted line labeled $L_{\rm BZ}=10^{40} \mbox{ ergs s}^{-1}$
cross at $M \sim 10^6 M_\odot$;
thus, we find that a super-massive BH with $M \sim 10^6 M_\odot$
has a gap whose luminosity can attain up to
$10^{40} \mbox{ ergs s}^{-1}$.

To consider the case of an efficient energy extraction from
a BH, we assume $a_\ast=0.9$ unless explicitly mentioned.
Substituting $\dot{m}=\dot{m}_{\rm up}$ into
equation~(\ref{eq:L_BZ_2}),
we then obtain the maximum gap luminosity 
\begin{equation}
  L_{\rm BZ} 
    \approx 5.2 \times 10^{35} a_\ast{}^2 M_1{}^{6/7} 
    \mbox{ergs s}^{-1}.
  \label{eq:L_BZ_3}
\end{equation}
Assuming that 100~\% of this power is converted into radiation,
we obtain the upper limit of its flux at Earth, 
$ F_{\rm BZ}= L_{\rm BZ} / 4\pi d^2$,
where $d$ is the distance to the BH.

For stellar-mass BHs, we obtain the flux upper limit,
\begin{equation}
  F_{\rm BZ}
    = \frac{L_{\rm BZ}}{4\pi d^2}
    = 4.1 \times 10^{-9} a_\ast{}^2 M_1{}^{6/7}
      \left(\frac{d}{\rm kpc}\right)^{-2}
    \mbox{ergs s}^{-1} \mbox{ cm}^{-2}.
  \label{eq:F_BZ}
\end{equation}
As will be seen in \S~\ref{sec:stellar},
the gap emission spectrum peaks between GeV and 10~GeV
for $M \sim 10 M_\odot$.
Thus, if a small portion of this BZ power (e.g., $0.1$~\%) 
is dissipated in the gap,
we can expect a large time-averaged high-energy (HE) flux
(e.g., $> 3 \times 10^{-12} \mbox{ ergs s}^{-1}$)
that is detectable with {\it Fermi}/LAT.
There is another component, 
which is formed by the inverse-Compton (IC) scatterings,
appears in VHE.
This component may be detectable
with ground-based, imaging atmospheric Cherenkov telescopes (IACT).
Note that equation~(\ref{eq:F_BZ}) 
merely gives the upper limit,
and that the actual photon flux can be obtained 
when we solve the gap electrodynamics
from the basic equations.
Will will examine this issue in \S~\ref{sec:stellar}.

For intermediate-mass BHs (IMBHs), we obtain the maximum flux of
\begin{equation}
  F_{\rm BZ}
    = 2.1 \times 10^{-9}
      a_\ast{}^2 M_3{}^{6/7}
      \left(\frac{d}{\rm 10~kpc}\right)^{-2}
    \mbox{ergs s}^{-1} \mbox{ cm}^{-2},
\end{equation}
where $M_3 \equiv M/(10^3 M_\odot)$.
The spectrum has two peaks:
the curvature photons peak in 1-10~GeV and the IC ones above TeV.
Both spectral components are potentially detectable in HE and VHE,
if an IMBH is located within our galaxy.
We will examine this possibility in \S~\ref{sec:IMBH}.

For supermassive BHs, we obtain the maximum flux of
\begin{equation}
  F_{\rm BZ}
    = 3.0 \times 10^{-10}
      a_\ast{}^2 M_9{}^{6/7}
      \left(\frac{d}{\rm 10~Mpc}\right)^{-2}
    \mbox{ergs s}^{-1} \mbox{ cm}^{-2},
\end{equation}
where $M_9 \equiv M/(10^9 M_\odot)$.
The IC component, which appears in VHE,
may be detectable with IACT.
We will examine this possibility in \S~\ref{sec:SMBH}.

\section{Background geometry and the null charge surface}
\label{sec:geometry}
The self-gravity of the plasma particles and 
the electromagnetic field little affects the space-time geometry.
Thus, around a rotating BH, 
the background geometry is described by the Kerr metric
\citep{kerr63}.
In the Boyer-Lindquist coordinates, it becomes 
\citep{boyer67} 
\begin{equation}
 ds^2= g_{tt} dt^2
      +2g_{t\varphi} dt d\varphi
      +g_{\varphi\varphi} d\varphi^2
      +g_{rr} dr^2
      +g_{\theta\theta} d\theta^2,
  \label{eq:metric}
\end{equation}
where 
\begin{equation}
   g_{tt} 
   \equiv 
   -\frac{\Delta-a^2\sin^2\theta}{\Sigma} c^2 ,
   \qquad
   g_{t\varphi}
   \equiv 
   -\frac{2(GM/c^2) ar \sin^2\theta}{\Sigma} c, 
  \label{eq:metric_2}
\end{equation}
\begin{equation}
   g_{\varphi\varphi}
     \equiv 
     \frac{A \sin^2\theta}{\Sigma} , 
     \qquad
   g_{rr}
     \equiv 
     \frac{\Sigma}{\Delta} , 
     \qquad
   g_{\theta\theta}
     \equiv 
     \Sigma ;
  \label{eq:metric_3}
\end{equation}
$\Delta \equiv r^2-2(GM/c^2)r+a^2$,
$\Sigma\equiv r^2 +a^2\cos^2\theta$,
$A \equiv (r^2+a^2)^2-\Delta a^2\sin^2\theta$.
At the event horizon, $\Delta$ vanishes, giving
$r_{\rm H} \equiv r_{\rm g}+\sqrt{r_{\rm g}{}^2-a^2}$
as the horizon radius. 
If the BH is extremely rotating 
(i.e., if $a \rightarrow r_{\rm g}$) for instance,
we obtain $r_{\rm H} \rightarrow r_{\rm g}$.

In a stationary and axisymmetric space time
(as described by eq.~\ref{eq:metric}),
the Gauss's law becomes
\begin{equation}
  \nabla_\mu F^{t\mu}
  = \frac{1}{\sqrt{-g}}
    \partial_\mu \left[ \frac{\sqrt{-g}}{\rhowSQR}
                       g^{\mu\nu}(-g_{\varphi\varphi}F_{t\nu}
                               +g_{t\varphi}F_{\varphi\nu})
               \right]
  = \frac{4\pi}{c^2} \rho,
  \label{eq:Poisson_1}
\end{equation}
where $\nabla$ denotes the covariant derivative,
the Greek indices run over $t$, $r$, $\theta$, $\varphi$;
$\sqrt{-g}= \sqrt{g_{rr}g_{\theta\theta}\rhowSQR}
 =c\Sigma\sin\theta$ and
$\rhowSQR \equiv g_{t\varphi}^2-g_{tt}g_{\varphi\varphi}=c^2 \Delta \sin^2\theta$,
$\rho$ the real charge density.
The electromagnetic fields are observed by an observer that is static
with respect to asymptotic infinity,
and are given by~\citep{came86a,came86b}
$ E_r=F_{rt}, \, E_\theta=F_{\theta t}, \, E_\varphi=F_{\varphi t}$,
$ B^r= (g_{tt}+g_{t\varphi}\Omega) F_{\theta\varphi}/\sqrt{-g}, \,
  B^\theta= (g_{tt}+g_{t\varphi}\Omega) F_{\varphi r}/\sqrt{-g}, \,
  B_\varphi= -\rhowSQR F_{r \theta}/\sqrt{-g}$,
where $F_{\mu\nu} \equiv \partial_\mu A_\nu-\partial_\nu A_\mu$ 
and $A_\mu$ denotes the vector potential.

In this paper, we assume that the electromagnetic fields 
(i.e., all components of the Faraday tensor, $F_{\mu\nu}$) 
depend on $t$ and $\varphi$ through 
$\varphi-\Omega_{\rm F} t$.
In this case, we can introduce the non-corotational potential $\Phi$
such that
\begin{equation}
  F_{\mu t}+\Omega_{\rm F} F_{\mu \varphi}
  = -\partial_\mu \Phi(r,\theta,\varphi-\Omega_{\rm F} t).
  \label{eq:def_Phi}
\end{equation}
If $F_{A t}+\Omega_{\rm F} F_{A \varphi}=0$ holds
for $A=r$ and $\theta$,
$\Omega_{\rm F} \equiv F_{tr}/F_{r\varphi}= F_{t\theta}/F_{\theta\varphi}$ 
is conserved along the field line.
However, in a particle acceleration region,
$F_{A t}+\Omega_{\rm F} F_{A \varphi}$ deviates from $0$
and the magnetic field does not rigidly rotate. 
The deviation from rigid rotation is expressed 
in terms of $\Phi$, which gives the strength of 
the acceleration electric field 
measured by an observer that is static to asymptotic infinity as
\begin{equation}
  \Ell \equiv \frac{\mbox{\boldmath$B$}}{B}
              \cdot \mbox{\boldmath$E$}
       = \frac{B^i}{B}(F_{it}+\Omega_{\rm F} F_{i\varphi})
       = \frac{\mbox{\boldmath$B$}}{B}
              \cdot (-\nabla\Phi),
  \label{eq:def_Ell}
\end{equation}
where the Latin index $i$ runs over spatial coordinates
$r$, $\theta$, $\varphi$.
Note that $B^i F_{i \varphi}=0$.

Substituting Eq.~(\ref{eq:def_Phi}) into (\ref{eq:Poisson_1}),
we obtain the Poisson equation for the
non-corotational potential,
\begin{equation}
  -\frac{c^2}{\sqrt{-g}}
   \partial_\mu 
      \left( \frac{\sqrt{-g}}{\rhowSQR}
             g^{\mu\nu} g_{\varphi\varphi}
             \partial_\nu \Phi
      \right)
  = 4\pi(\rho-\rhoGJ),
  \label{eq:pois}
\end{equation}
where the GR Goldreich-Julian charge density
is defined as
\begin{equation}
  \rhoGJ \equiv 
      \frac{c^2}{4\pi\sqrt{-g}}
      \partial_\mu \left[ \frac{\sqrt{-g}}{\rhowSQR}
                         g^{\mu\nu} g_{\varphi\varphi}
                         (\Omega_{\rm F}-\omega) F_{\varphi\nu}
                 \right].
  \label{eq:def_GJ}
\end{equation}
In the limit $r \gg r_{\rm g}$, 
equation~(\ref{eq:def_GJ}) reduces to the
ordinary, special-relativistic expression~\citep{GJ69,mestel71},
\begin{equation}
  \rhoGJ 
  \equiv -\frac{\mbox{\boldmath$\Omega$}\cdot\mbox{\boldmath$B$}}
               {2\pi c}
         +\frac{(\mbox{\boldmath$\Omega$}\times\mbox{\boldmath$r$})\cdot
                (\nabla\times\mbox{\boldmath$B$})}
               {4\pi c}.
  \label{eq:def_rhoGJ_1}
\end{equation}
Therefore, the corrections due to magnetospheric currents,
which are expressed by the second term of eq.~(\ref{eq:def_rhoGJ_1}),
are included in equation~(\ref{eq:def_GJ}).


In addition to $F_{\mu\nu}$ and $\Phi$,
$\rho$ may also depend on $t$ and $\varphi$ 
only through $\varphi-\Omega_{\rm F} t$.
In this case, equation~(\ref{eq:pois}) gives a \lq stationary' 
gap solution in the \lq co-rotational' frame,
in the sense that $\Phi$ and $\rho$ are a function of 
$r$, $\theta$, and $\varphi-\Omega_{\rm F} t$.
Note that such stationary solutions are valid not only between the
two light surfaces (i.e., where 
$k_0 \equiv -g_{tt}-2g_{t\varphi}\Omega_{\rm F}
           -g_{\varphi\varphi}\Omega_{\rm F}^2 > 0$),
but also inside the inner light surface and outside the outer
light surface (i.e., where $k_0<0$).

Equation~(\ref{eq:pois}) shows that
$\Ell$ is exerted along $\mbox{\boldmath$B$}$
if $\rho$ deviates from $\rhoGJ$ in any region.
In the close vicinity of the inner and outer boundaries,
we can neglect the trans-field derivatives to obtain
$\partial_r E_\parallel \propto \rho-\rhoGJ$.
Thus, $\partial_r E_\parallel$, and hence $\rho-\rhoGJ$ 
must have different signs at the two boundaries
so that the gap may be closed.
In a nearly vacuum gap, $\rho \approx 0$ shows that
$\rhoGJ$ should change sign within the gap.
Therefore, a gap should appear around the null-charge surface,
where $\rhoGJ$ vanishes.
The null surface is, therefore,
a natural place for a particle accelerator (i.e., a gap) to arise,
in the same way as pulsar vacuum gap models
\citep{hollo73,chiang92,romani96,cheng00}. 

It should be noted that
the null surface appears
{\bf near} the place where $\Omega_{\rm F}$ coincides with 
the space-time dragging angular frequency, $\omega$ \citep{bes92}.
The deviation of the null surface
from this $\omega(r,\theta)=\Omega_{\rm F}$ surface is,
indeed, small, as figure~1 of \citet{hiro98} indicates.
Since $\omega$ matches $\Omega_{\rm F}$ only near the horizon,
the null surface, and hence the gap generally appears 
within one or two gravitational radii above the horizon,
irrespective of the BH mass.

\section{Magnetospheric lepton accelerator near the horizon}
\label{sec:accelerator}
In this section, we formulate the BH-gap electrodynamics,
extending the method described in \citet{hiro13} and HP16.
Throughout this paper, we assume 
an aligned rotator in the sense that the magnetic axis
coincides with the rotational axis of the BH,
and seek an axisymmetric solution.

\subsection{Magnetic field structure}
\label{sec:mag}
As described in HP16, a stationary BH gap is formed 
around the null surface,
as long as the injected current density
across the inner or outer boundaries is much small compared 
to the GJ value.
Since the null surface is formed by the frame-dragging effect,
the gap electrodynamics is essentially
governed by the frame-dragging effect
rather than the magnetic field configurations.
This forms a striking contrast to 
the pulsar outer-magnetospheric gap model,
in which the null surface is formed 
by the convex geometry of the poloidal magnetic field lines.
Thus, in a BH magnetosphere, 
the gap position and its spatial extent,
as well as the exerted $E_\parallel=E_\parallel(s)$ in the gap,
little depend on the magnetic flux function,
$\Psi=A_\varphi$.
We thus assume a radial magnetic field on the poloidal plane,
$\Psi=\Psi(\theta)$.

Because of axial symmetry, the gap electrodynamic structure
can be described in the poloidal plane.
For simplicity, we assume that the photons propagate radially
in this 2-D plane,
which is justified if the photons have negligible angular momenta.
Provided that the drift motion 
(e.g., due to toroidal radiation drag in a radial magnetic field, 
or due to radial gravity in a toroidal magnetic field) 
is small in the meridional direction,
charged particles roughly move along the magnetic field lines
in the poloidal plane.
In this case, due to relativistic beaming, ultra-relativistic particles
emit photons along the instantaneous magnetic field lines;
thus, as long as the poloidal magnetic field is radial,
photons propagate on the same magnetic flux surface,
$A_\varphi=A_\varphi(\theta)={\rm constant}$.
As a result, we can solve the particle equations of motion
and the radiative transfer equation along individual 
radial poloidal magnetic field lines separately.

Even when the magnetic field lines are radial in the poloidal plane,
there exists a toroidal magnetic component, $B_\varphi$,
due to retardation, magnetospheric currents, and a frame dragging.
It is, however, out of the scope of this paper
to restrict the functional form of $B_\varphi(r,\theta)$,
taking account of such effects.
In addition, $B_\varphi$ does not affect $\rhoGJ$ or
the propagation direction of particles and photons
in the poloidal plane.
Thus, we do not specify $B_\varphi$.
Accordingly, instead of computing the curvature radius, $R_{\rm c}$,
of the leptons from their 3-D motion in the rotating magnetosphere,
we parameterize $R_{\rm c}$ when we calculate the curvature emission.
Although $R_{\rm c}$ affects the spectral shape of curvature emission,
it little affects the total luminosity,
because the latter is essentially determined by the potential drop 
within the gap, and because the potential drop is determined by the
gap width, 
which is in turn predominantly determined by the pair production rate of
the IC-emitted, VHE photons 
(not the curvature-emitted, lower energy photons).
On these grounds, we adopt $R_{\rm c}=r_{\rm g}$
in the present paper, leaving $B_\varphi$ unconstrained.


\subsection{Gap electrodynamics}
\label{sec:ED}
In the same way as HP16,
we solve the stationary gap solution
from the set of the Poisson equation for $\Phi$,
the equations of motion for electrons and positrons,
and the radiative transfer equation for the emitted photons.

\subsubsection{Poisson equation}
\label{sec:pois}
\begin{figure}
 \includegraphics[angle=-90,scale=0.45]{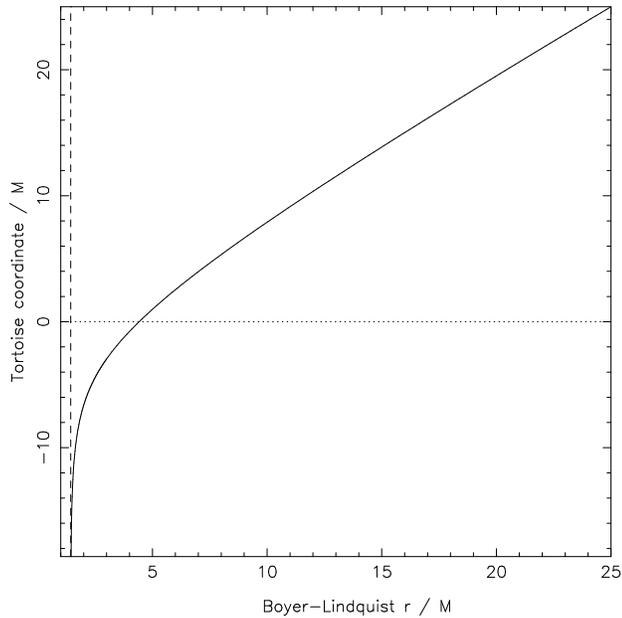}
\caption{
Tortoise coordinate versus Boyer-Lindquist radial coordinate.
Both axes are normalized by the gravitational radius, $r_{\rm g}$.
The BH spin parameter is adopted to be $a_\ast=0.90$. 
Vertical dashed line represents the horizon position.
\label{fig:tortois}
}
\end{figure}

To solve the radial dependence of $\Phi$
in the Poisson equation~(\ref{eq:pois}),
we introduce the following dimensionless 
tortoise coordinate, $\eta_\ast$,
\begin{equation}
  \frac{d\eta_\ast}{dr}= \frac{r^2+a^2}{\Delta} \frac{1}{r_{\rm g}}.
  \label{eq:tortois}
\end{equation}
In this coordinate, the horizon corresponds to 
the \lq inward infinity', $\eta_\ast=-\infty$.
In this paper, we set $\eta_\ast=r/r_{\rm g}$ 
at $r=r_{\rm max} \equiv 25 r_{\rm g}$,
where the value of $r_{\rm max}$ can be chosen arbitrarily
and does not affect the results in any ways.
The distribution of $\eta_\ast$ is depicted as a function of $r/r_{\rm g}$
in figure~\ref{fig:tortois}.
Note that the relationship between $\eta_\ast$ and $r$ 
does not depend on the colatitude, $\theta$.

Since the gap is located near the horizon, 
we take the limit $\Delta \ll r_{\rm g}{}^2$.
Assuming that $\Phi$ does not depend on $\varphi-\Omega_{\rm F} t$,
that is, $\Phi=\Phi(r,\theta)$,
we can recast the Poisson equation~(\ref{eq:pois})
into the two-dimensional form,
\begin{eqnarray}
  &&
  -\left(\frac{r^2+a^2}{\Delta}\right)^2
   \frac{\partial^2 \tilde{\Phi}}{\partial \eta_\ast{}^2}
  +\frac{2(r-r_{\rm g})(r^2+a^2)}{\Delta^2/r_{\rm g}}
   \frac{\partial \tilde{\Phi}}{\partial \eta_\ast}
   \nonumber\\
  &&
  -\frac{r_{\rm g}{}^2}{\Delta}
   \frac{\Sigma}{\sin\theta}
   \frac{\partial}{\partial\theta}
   \left( \frac{\sin\theta}{\Sigma}
          \frac{\partial\tilde{\Phi}}{\partial\theta} 
   \right)
   \nonumber\\
  &&
   = \left(\frac{\Sigma}{r^2+a^2}\right)^2 
     (n_+ -n_- -n_{\rm GJ}),
  \nonumber\\
  \label{eq:pois_3}
\end{eqnarray}
where 
\begin{equation}
  \tilde{\Phi}(\eta_\ast) 
  \equiv \frac{c}{2\Omega_{\rm F}B r_{\rm g}{}^2} \Phi(r)
  \label{eq:def_Phi2}
\end{equation}
denotes the dimensionless non-corotational potential.
Dimensionless lepton densities per magnetic flux tube
are defined by
\begin{equation}
  n_\pm \equiv \frac{2\pi ce}{\Omega_{\rm F} B} N_\pm,
  \label{eq:def_n}
\end{equation}
where the number densities of positrons and electrons,
$N_+$ and $N_-$, are computed from the pair production rate
at each position
\citep{hiro98,hiro99a,hiro99b}.
Dimensionless GJ charge density per magnetic flux tube
is defined by
\begin{equation}
  n_{\rm GJ} \equiv \frac{2\pi c}{\Omega_{\rm F} B} \rhoGJ .
  \label{eq:def_nGJ}
\end{equation}

For a radial poloidal magnetic field, $\Psi=\Psi(\theta)$,
we can compute the acceleration electric field by
\begin{equation} 
  E_\parallel 
    \equiv -\frac{\partial \Phi}{\partial r}
    = -r_{\rm g} \frac{\Omega_{\rm F} B}{c}
      \frac{r^2+a^2}{\Delta}
      \frac{\partial \tilde\Phi}{\partial \eta_\ast}.
  \label{eq:Ell2}
\end{equation}
Without loss of any generality, 
we can assume $F_{\theta\varphi}>0$ in the northern hemisphere.
In this case, a negative $E_\parallel$ arises in the gap,
which is consistent with the direction 
of the global current flow pattern. 

Equation~(\ref{eq:def_GJ}) shows that $\rhoGJ$ 
is essentially determined by $B^r$, rather than $B^\theta$,
near the horizon.
In a stationary and axisymmetric magnetosphere,
equation~(\ref{eq:def_GJ}) becomes
\begin{equation}
  \rhoGJ 
  \propto \partial_r (G F_{\varphi r}) 
         +\frac{1}{\Delta} \partial_\theta(G F_{\varphi\theta}),
  \label{eq:rhoGJ_2} 
\end{equation}
where $G \equiv (A\sin\theta/\Sigma)(\Omega_{\rm F}-\omega)$.
Since $G$, $F_{\varphi r}$, and $F_{\varphi\theta}$
are well-behaved at the horizon,
we find that the second term dominates the first one 
at $\Delta \rightarrow 0$.
Therefore, although the null surface itself is formed by the 
{\it frame-dragging} effect, the radial component of the 
magnetic field essentially determines $\rhoGJ$
near the horizon owing to the {\it redshift} effect.

\subsubsection{Leptonic densities}
\label{sec:densities}
We next consider $n_-$ and $n_+$ in equation~(\ref{eq:pois_3}).
Because of $E_\parallel<0$,
electrons are accelerated outwards, while positrons inwards. 
As s result, as long as there is no current injection
across either outer or inner boundaries,
charge density, $n_+-n_-$, becomes negative (or positive)
at the outer (or inner) boundary.
In a stationary gap, $E_\parallel$ should not change sign in it.
In a vacuum gap, a positive (or a negative) $-n_{\rm GJ}$
near the outer (or inner) boundary makes 
$\partial_r E_\parallel >0 $ (or $\partial_r E_\parallel <0$),
thereby closing the gap.
In a non-vacuum gap, the right-hand side of equation~(\ref{eq:pois_3})
should become positive (or negative) near the outer (or inner)
boundary so that the gap may be closed.
Therefore, $\vert n_+-n_- \vert$ should not exceed 
$\vert n_{\rm GJ} \vert$ at either boundary.
At the outer boundary, for instance, we can put
\begin{equation}
  (n_+-n_-)\vert_{r=r_2} = -n_{-}(r=r_2) 
  = j n_{\rm GJ}(r=r_2),
\end{equation}
where the dimensionless parameter $j$ should be
in the range, $0 \le j \le 1$, so that the gap solution may be stationary.
Since $F_{\theta\varphi}>0$ is assumed, 
it is enough to consider a positive $j$.
If $j=1$, there is no surface charge at the outer boundary.
However, if $j<1$, the surface charge results in 
a jump of $\partial_r E_\parallel$ at the outer boundary.
That is, the parameter $j$ specifies the strength of 
$\partial_r E_\parallel$ at the outer boundary.
Thus, the inner boundary position, $r=r_1$, 
is determined as a free boundary
problem by this additional constraint, $j$.
The outer boundary position, $r=r_2$, or equivalently the gap width
$w= r_2 -r_1$, is constrained by the gap closure condition
(\S~\ref{sec:closure}).

It is noteworthy that the charge conservation ensures that
the dimensionless current density (per magnetic flux tube),
$J_{\rm c} \equiv -n_+ -n_-$ 
conserves along the flowline.
At the outer boundary, we obtain
\begin{equation}
  J_{\rm c}= -n_-(r=r_2)= j n_{\rm GJ}(r=r_2).
  \label{eq:J_outBD}
\end{equation}
Thus, $j$ specifies not only $\partial_r E_\parallel$ at the
outer boundary, but also the conserved current density, $J_{\rm c}$.

In general, under a given electro-motive force exerted 
in the ergosphere,
$J_{\rm c}$ should be constrained by the global current 
flow pattern,
which includes an electric load at the large distances where the
force-free approximation breaks down and the trans-magnetic-field
current gives rise to the outward acceleration of charged particles 
by Lorentz forces (thereby converting the Poynting flux into
particle kinetic energies).
However, we will not go deep into the determination of $J_{\rm c}$
in this paper,
because we are concerned with the acceleration processes
near the horizon,
not the global current closure issue. 
Note that $w$ (or $r_2$) is essentially determined by $\dot{m}$;
thus, $j$ and $\dot{m}$ give the actual current density 
$(\Omega_{\rm F} B / 2\pi) J_{\rm c}$,
where $B$ should be evaluated at each position.
On these grounds, instead of determining $J_{\rm c}$ 
by a global requirement,
we treat $j$ as a free parameter in the present paper.

It may be worth mentioning, in passing, that 
we may not have to consider a time-dependent solution,
which may be obtained when $j > 1$ 
as in pulsar polar-cap models
\citep{harding78,daugherty82,dermer94}.
In the polar cap model, the absence of the null surface
results in a non-stationary gap solution
\citep{timo10,timo13,timo15}.
However, in the pulsar outer gap models 
\citep{zhang97,romani10,takata06,wangY11,hiro15}
or in the present BH gap model,
the existence of the null surface leads to a formation
of a stationary gap around this surface.
We thus adopt $0 \le j \le 1$ and consider a stationary gap solution.


\subsubsection{Particle motion}
\label{sec:EOM}
Let us describe the motion of electrons and positrons.
For simplicity, we assume that the distribution functions
of electrons and positrons are mono-energetic.
We evaluate the Lorentz factors of electrons (or positron) 
by the motion of a test particle injected
across the inner (or the outer) boundary of the gap.
For example, an injected test electron is accelerated
by a negative $E_\parallel$ outwards and 
loses the kinetic energy via curvature and IC processes.
The former, curvature radiation rate (per particle)
is computed by the standard
synchrotron emission formula with the gyration radius
replaced with $R_{\rm c}$
\citep{rybicki79}.
The latter, IC radiation rate 
is computed by multiplying the
scattering probability (per unit time) 
and the scattered photon energy.
Thus, the particle Lorentz factor saturates at the curvature-
or IC-limited value, whichever smaller.

\subsubsection{Radiative transfer equation}
\label{sec:RTE}
Throughout this paper, we assume that all photons are emitted
with vanishing angular momenta.
In this case, photons propagate on a constant-$\theta$ surface; thus,
the radiative transfer equation is solved one-dimensionally
along the radial magnetic field lines on the poloidal plane.
These primary leptons emit photons via curvature and IC processes
both inside and outside the gap.
For the details of how to compute the emissivities of 
curvature and IC processes, see \S\S~4.2 and 4.3 of HP16. 

Some portions of the photons are emitted above 10~TeV via IC process.
A significant fraction of such hard $\gamma$-rays
are absorbed, colliding with the RIAF soft photons.
If such collisions take place within the gap, 
the created electrons and positrons polarize 
to be accelerated in opposite directions,
becoming the primary leptons.
If the collisions take place outside the gap,
the created, secondary pairs migrate along the magnetic filed lines
to emit photons via IC and synchrotron processes.
Some of such secondary IC photons are absorbed again
to materialize as tertiary pairs, 
which emit tertiary photons via synchrotron and IC processes,
eventually cascading into higher generations.
As a representative model of the RIAF,
we adopt the analytic solution of the 
advection-dominated accretion flow (ADAF) obtained by
\citet{mahad97}.

We assume that the ADAF soft photon specific intensity
is isotropic in the zero-angular-momentum observer (ZAMO).
This assumption simplifies the calculations of 
photon-photon pair production and IC scatterings,
because photons are assumed to be emitted by leptons
with vanishing angular momenta.
To calculate the flux of ADAF photons,
we assume that their number density is homogeneous and becomes
$L_{\rm ADAF}/(4\pi R_{\rm min}^2 c)$ within $r<R_{\rm min}$, where 
$L_{\rm ADAF}$ denotes the ADAF luminosity given by \citet{mahad97}.
We assume that the ADAF luminosity becomes $L_{\rm ADAF}$
at $r=R_{\rm min}$.
Outside this radius, $r>R_{\rm min}$, 
we assume that their number density decreases by $r^{-2}$ law. 
This treatment may be justified, because the submillimeter-IR photons,
which most effectively work both for pair production and IC scatterings,
are emitted from the inner-most region of the ADAF.

\subsubsection{Boundary conditions}
\label{sec:BDCs}
We solve the gap in the 2-D poloidal plane.
We assume a reflection symmetry with respect to the magnetic axis.
Thus, we put $\partial_\theta \tilde\Phi=0$ at $\theta=0$.
We assume that the polar funnel is bounded at
a fixed colatitude, $\theta=\theta_{\rm max}$
and impose that this lower-latitude boundary
is equi-potential and put $\tilde\Phi=0$ at $\theta=\theta_{\rm max}$.

Both the outer and inner boundaries are treated as free
boundaries.
Their positions are determined by the two conditions,
the value of $j$ along each magnetic field line
(specified by $\Psi$), and
the gap closure condition 
(to be described in \S~\ref{sec:closure}).
For simplicity, we assume that $j$ is constant for $\Psi$.
At the outer boundary, 
$\partial_r E_\parallel= -\partial_r{}^2 \Phi$
is specified by $j$.
At the inner boundary, we impose
$E_\parallel= -\partial_r \Phi=0$.
We assume that electrons, positrons or photons
are not injected across either the outer or the inner 
boundaries.

\subsubsection{Gap closure condition}
\label{sec:closure}
The set of Poisson and radiative-transfer equations
are solved together with the terminal Lorentz factor $\gamma$
and the $n_\pm$ obtained by the local pair production rate. 
Unlike HP16, we discard the reflection symmetry 
(along radial magnetic field lines) 
with respect to the null-charge surface, 
and explicitly consider the asymmetric 
distribution of $E_\parallel$, 
$\gamma$ and $n_\pm$, and the photon specific intensity 
at each point $r$.
Accordingly, the gap closure condition should be modified as
${\cal M}_{\rm in} {\cal M}_{\rm out} = 1$, where
${\cal M}_{\rm in}$ and ${\cal M}_{\rm out}$ denote the multiplicity
(eq.~[41] of HP16)
associated with the in-going and out-going leptons, respectively.
See \citet{hiro13} for a detailed treatment of the asymmetric
multiplicities, $\cal{M}_{\rm in} \ne \cal{M}_{\rm out}$.

\section{Gap solutions}
\label{sec:solutions}
In \S~\ref{sec:detect},
we examined the upper limits of the BH gap luminosity, 
imposing the charge-starvation condition (eq.~[\ref{eq:mdot_up}])
that requires the ADAF to be less luminous so that their
MeV photons may not produce pairs above the GJ density.
In this section, solving the gap electrodynamics
by the method described in \S~\ref{sec:accelerator}
for various BH masses,
we demonstrate that the gap becomes most luminous 
when the gap longitudinal width becomes much greater than
the horizon radius,
and that there is a lower-limit accretion rate
below which a stationary pair-production cascade cannot be maintained 
within the gap.

Throughout this paper, 
we assume a radial poloidal magnetic field,
$\Psi=\Psi(\theta)\propto -\cos\theta$.
The magnitude of $\Psi$ is adjusted so that $B=B_{\rm eq}$
(eq.~[\ref{eq:B_eq}]) may be satisfied at $r=2r_{\rm g}$.
Assuming that the poloidal magnetic components dominate 
the toroidal one, we put
$B(r)=B_{\rm eq}(r/2r_{\rm g})^{-2}$
and compute the synchrotron emission outside the gap.
Unless explicitly mentioned,
we adopt $a=0.90 r_{\rm g}$,
$\Omega_{\rm F}=0.50 \omega_{\rm H}$, 
$R_{\rm min}=6 r_{\rm g}$, $j=0.7$, and
$R_{\rm c}=r_{\rm g}$.
To solve the Poisson equation~(\ref{eq:pois_3}),
we set the meridional boundary at $\theta=\theta_{\rm max}=60^\circ$.
We assume that the entire system is axisymmetric with respect
to the rotation axis, 
which coincides with the magnetic axis.

\subsection{The case of stellar-mass BHs}
\label{sec:stellar}
Let us first examine the case of $M=10 M_\odot$.
We begin with describing the 2-D distribution of the
$E_\parallel$ in \S~\ref{sec:stellar_e2d},
and demonstrate that the gap emission becomes strongest
within the colatitude,
$\theta < 38^\circ$ along the magnetic axis 
in \S~\ref{sec:stellar_SED_theta}.
Then in \S~\ref{sec:stellar_density}--\S~\ref{sec:stellar_Ell},
we examine the lepton densities and acceleration electric field.
Adopting a distance of $1$~kpc,
we present the expected spectra of gap emissions
in \S~\ref{sec:stellar_SED},
and examine its dependence on $j$, $a_\ast$, and $\Omega_{\rm F}$
in \S\S~\ref{sec:stellar_current}--\ref{sec:stellar_omegaF}.
Finally, in \S~\ref{sec:stellar_cascaded},
we demonstrate that the magnetosphere becomes
entirely force-free (except for the gap region)
by the cascaded pairs outside the gap.

\subsubsection{Electric field along the magnetic field lines}
\label{sec:stellar_e2d}
\begin{figure}
  \includegraphics[angle=0,scale=0.70]{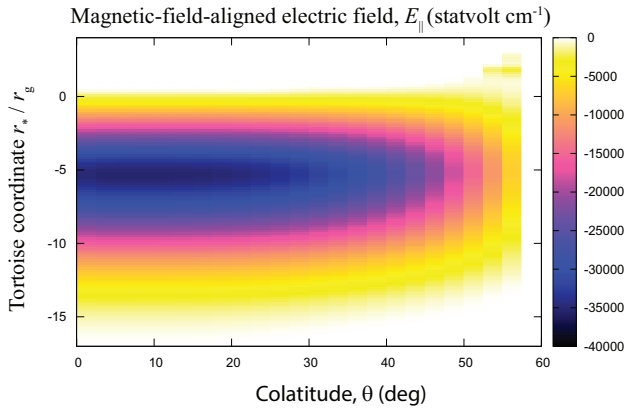}
\caption{
Acceleration electric field (statvolt ${\rm cm}^{-1}$)
on the poloidal plane.
The abscissa denotes the magnetic colatitudes, $\theta$,
in degrees,
where $0$ (i.e., the ordinate) corresponds to the magnetic axis.
The ordinate denotes the dimensionless tortoise coordinate,
where $-\infty$ corresponds to the event horizon.
The gap is solved for a black hole with $M=10 M_\odot$
and $a_\ast=0.9$.
Dimensionless accretion rate is chosen to be
$\dot{m}=1.00 \times 10^{-4}$.
\label{fig:e2d_1e1}
}
 \end{figure}

We first present the distribution of the
magnetic-field-aligned electric field on the poloidal plane.
In figure~\ref{fig:e2d_1e1},
we plot $E_\parallel$ (in statvolt ${\rm cm}^{-1}$)
as a function of
the dimensionless tortoise coordinate, $\eta_\ast$, and
the magnetic colatitude, $\theta$ (in degrees),
for $\dot{m}= 1.00 \times 10^{-4}$.

\begin{figure}
  \includegraphics[angle=0,scale=0.40]{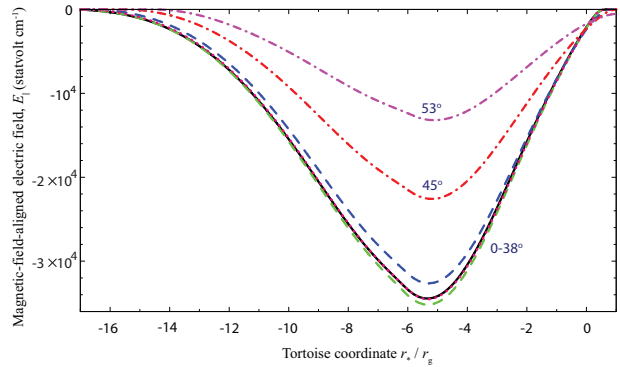}
\caption{
Acceleration electric field at six discrete colatitudes, $\theta$,
as a function of the dimensionless tortoise coordinate,
$\eta_\ast$, for $\dot{m}=1.00 \times 10^{-4}$.
The values of $E_\parallel(\eta_\ast,\theta)$ in
figure~\ref{fig:e2d_1e1} is plotted at six discrete $\theta$'s:
the black solid, green dashed, blue dashed,
magenta dotted, red dash-dotted, and purple dash-dotted
curves denote the $E_\parallel$
at $\theta= 0^\circ$, $15^\circ$, $30^\circ$, $37.5^\circ$,
$45^\circ$, and $52.5^\circ$, respectively. 
\label{fig:e2d_th_1e1}
}
 \end{figure}

We also plot $E_\parallel$ at six discrete colatitudes
in figure~\ref{fig:e2d_th_1e1}.
It follows that the $E_\parallel(\eta_\ast)$ distribution
little changes in the polar region within
$\theta<38^\circ$.

\subsubsection{Gap emission versus colatitudes}
\label{sec:stellar_SED_theta}
\begin{figure}
  \includegraphics[angle=0,scale=0.40]{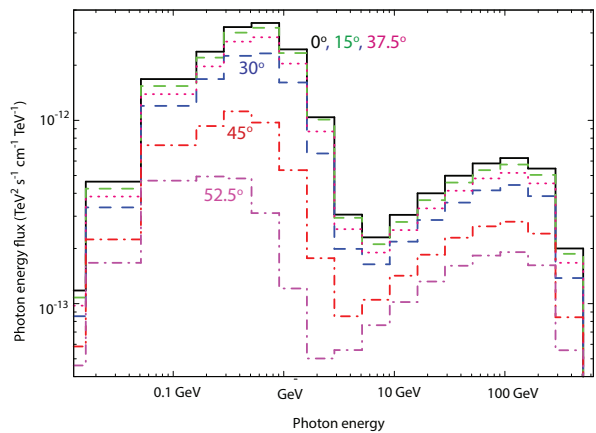}
\caption{
Spectral energy distribution (SED) of the emission from 
a black hole (BH) gap at six discrete colatitudes, $\theta$,
at a distance of 1~kpc.
The gap is solved for a black hole with $M=10 M_\odot$
and $a_\ast=0.9$.
The dimensionless accretion rate is fixed at 
$\dot{m}=1.00 \times 10^{-4}$.
The six lines correspond to the same $\theta$'s as in 
figure~\ref{fig:e2d_th_1e1}.
\label{fig:SED_th_1e1}
}
 \end{figure}

We next compare the $\gamma$-ray spectra
of a BH gap emission
as a function of the colatitude, $\theta$.
In figure~\ref{fig:SED_th_1e1},
we compare the SEDs at the same six discrete $\theta$'s
as in figure~\ref{fig:e2d_th_1e1}.
It follows that the gap emission becomes
most luminous if we observe the gap
with a viewing angle $\theta < 38^\circ$.
This conclusion is unchanged if we adopt different BH masses or spins.
In what follows, we therefore adopt
$\theta=0$ as the representative colatitude
to estimate the maximum $\gamma$-ray flux
of BH gaps.

\subsubsection{Created lepton densities}
\label{sec:stellar_density}
\begin{figure}
  \includegraphics[angle=0,scale=0.40]{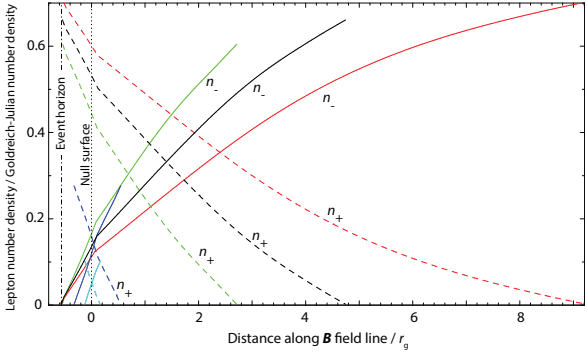}
\caption{
Spatial distribution of the electronic (solid) and
positronic (dashed) densities per magnetic flux tube
as a function of the distance, $r-r_0$, from the 
null-charge surface, in the Boyer-Lindquist radial coordinate
(in $r_{\rm g} \equiv GMc^{-2}$ unit).
The gap is solved for a black hole with $M=10 M_\odot$
and $a_\ast=0.9$.
Viewing angle is chosen to be $\theta=0$.
The cyan, blue, green, black, and red
curves represent the solutions for
$\dot{m}=1.00 \times 10^{-3}$,
$3.16 \times 10^{-4}$,
$1.00 \times 10^{-4}$,
$7.49 \times 10^{-5}$, and 
$5.62 \times 10^{-5}$, respectively.
The vertical dashed line at $r-r_0=0$ 
shows the null surface position.
\label{fig:crtd_1e1}
}
 \end{figure}

We plot the solved lepton densities at five discrete $\dot{m}$
in figure~\ref{fig:crtd_1e1}.
Since $E_\parallel$ is negative,
electrons are accelerated outwards while positrons inwards.
Thus, the dimensionless electronic density (solid curve), $n_-$, 
per magnetic flux tube, increases outwards,
while the positronic one (dashed curve), $n_+$, decreases outwards.  
Note that the abscissa, $r-r_0$,
is converted from the tortoise coordinate
to the Boyer-Lindquist radial coordinate for presentation purpose.
Thus, $r-r_0=0$ corresponds to the null-charge surface.
These solved $n_\pm(r,\theta)$ are used to compute
the real charge density $\rho=(\Omega_{\rm F}B/2\pi c)(n_+-n_-)$ 
at each position,
which is necessary to solve $E_\parallel$ 
on the poloidal plane.
We continue iterations until $n_\pm(\eta_\ast,\theta)$, 
$E_\parallel(\eta_\ast,\theta)$,
and the photon specific intensity saturate.

\subsubsection{Acceleration electric field}
\label{sec:stellar_Ell}
\begin{figure}
  \includegraphics[angle=0,scale=0.40]{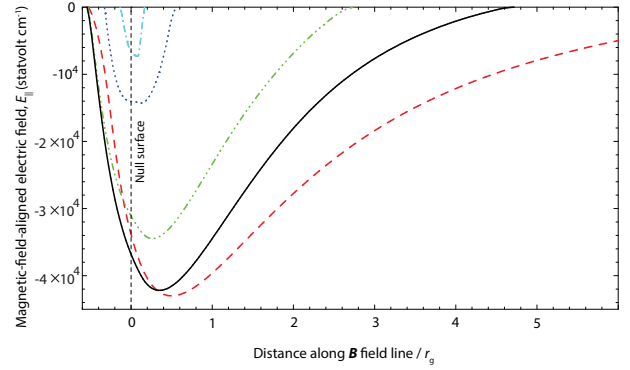}
\caption{
Spatial distribution of the magnetic-field-aligned
electric field, $E_\parallel$,
for five discrete accretion rates.
The gap is solved for a black hole with $M=10 M_\odot$
and $a_\ast=0.9$.
Each color corresponds to the same cases of $\dot{m}$'s
as in figure~\ref{fig:crtd_1e1}.
The vertical dashed line shows the null surface position.
\label{fig:Ell_1e1}
}
 \end{figure}

As $\dot{m}$ decreases, the reduced ADAF near-IR photon field
leads to less efficient pair production,
thereby resulting in an extended gap to sustain the externally
required current density, $j$, per magnetic flux tube.
We plot $E_\parallel(r,\theta=0)$ for five discrete $\dot{m}$'s
in figure~\ref{fig:Ell_1e1}.
The cyan, blue, green, black, and red curves
correspond to the cases of
$\dot{m}=10^{-3.0}$, $10^{-3.5}$, $10^{-4.0}$, $10^{-4.125}$, and 
$10^{-4.25}$, respectively;
that is, same as figure~\ref{fig:crtd_1e1}.
Integrating $E_\parallel$ over the gap width, 
we obtain the potential drop at each $\dot{m}$.
It becomes 
$-6.3 \times 10^{11}$~V,
$-3.9 \times 10^{12}$~V,
$-2.5 \times 10^{13}$~V,
$-4.2 \times 10^{13}$~V, and
$-6.3 \times 10^{13}$~V,
for 
$\dot{m}=10^{-3.0}$, $10^{-3.5}$, $10^{-4.0}$, $10^{-4.125}$, and 
$10^{-4.25}$, respectively.
Thus, the potential drop 
increases with decreasing $\dot{m}$
because of the increased gap width, $w \equiv r_2 -r_1$.
More specifically, as the accretion rate reduces, the decreased ADAF
near-IR photon field results in a less effective pair production
for the gap-emitted IC photons, 
thereby increasing the mean-free path for two-photon collisions.
Since $w$ essentially becomes
the pair-production mean-free path divided by the
number of photons emitted by a single electron
above the pair production threshold energy \citep{hiro98},
the reduced pair production leads to an extended gap 
along the magnetic field lines.
As a result, the smaller $\dot{m}$ is,
the greater the potential drop becomes.

As $w$ increases, the trans-field derivative 
begins to contribute in the Poisson equation~(\ref{eq:pois_3}). 
As a result, the $E_\parallel$ distribution shifts outwards,
in the same way as in pulsar outer-magnetospheric gaps
\citep[fig.~12 of][]{hiro99a}.
That is, a pulsar outer gap extends
from the null surface to (or beyond) the light cylinder
because the transverse thickness is limited by
the efficient screening of $E_\parallel$ due to the
trans-field propagation in concave poloidal magnetic field lines,
whereas a BH gap shifts from the null surface
towards the outer light surface
because the longitudinal width becomes comparable to the 
transverse thickness when the accretion rate is small.

\begin{figure}
  \includegraphics[angle=0,scale=0.40]{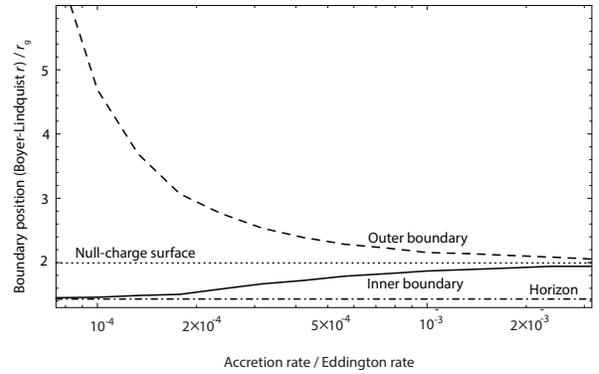}
\caption{
Spatial extent of the gap along the radial poloidal
magnetic field line
as a function of the dimensionless accretion rate.
The gap is solved for a black hole with 
$M=10 M_\odot$ and $a_\ast=0.9$.
Thick solid and dashed curves denote the position
of the inner and the outer boundaries of the gap,
in the unit of the gravitational radius,
$r_{\rm g} \equiv GMc^{-2}$. 
These boundary positions (in the ordinate) 
are transformed from the tortoise coordinate
into the Boyer-Lindquist coordinate.
The horizontal dash-dotted line shows the horizon radius,
whereas the horizontal dotted line does the null surface position
on the magnetic axis, $\theta=0$.
\label{fig:width_1e1}
}
 \end{figure}

Let us briefly examine how the gap width, $w$, is affected 
when the ADAF soft photon field changes.
In figure~\ref{fig:width_1e1}, we plot 
the gap inner and outer boundary positions 
as a function of $\dot{m}$,
where the ordinate is converted into 
the Boyer-Lindquist radial coordinate.
It follows that the gap inner boundary (solid curve, $r=r_1$), 
infinitesimally approaches the horizon
(dash-dotted horizontal line, $r=r_{\rm H}$),
while the outer boundary (dashed curve, $r=r_2$) moves outwards, 
with decreasing $\dot{m}$.
Below the accretion rate 
$\dot{m} \sim \dot{m}_{\rm low}= 2 \times 10^{-4}$,
the outer boundary moves rapidly away 
from the horizon with decreasing $\dot{m}$,
so that the required current density, $j=0.7$, 
may be produced within the gap 
under a diminished ADAF photon field.
At $\dot{m}=5.62 \times 10^{-5}$, and $4.21 \times 10^{-5}$,
the outer boundary
is located at $r_2=9.35 r_{\rm g}$ and $28.64 r_{\rm g}$.
We consider that the solution with $r_2 > 10 r_{\rm g}$
may not have significant physical meaning,
because the funnel boundary with the equatorial disk
will deviate from a conical shape beyond this radius
\citep{mckinney04,hirose04,krolik05,mckinney06,mckinney12,oriordan16}.
We thus define the lower-limit accretion rate,
$\dot{m}_{\rm low}$ when $r_2$ exceeds $10 r_{\rm g}$ in this paper.
Indeed, at further lower accretion rate, 
$\dot{m}\le 3.16 \times 10^{-5}$,
we fail to find a 2-D gap solution,
because the weak ADAF photon field can no longer sustain
the current density, $j=0.7$, per magnetic flux tube.
Thus, we obtain $\dot{m}_{\rm low} = 5.56 \times 10^{-5}$
for $M=10 M_\odot$, $a=0.90 r_{\rm g}$,
$\Omega_{\rm F}=0.50 \omega_{\rm H}$, and $j=0.7$,
interpolating $r_2=r_2(\dot{m})$ and putting $r_2=10r_{\rm g}$.
Because $r_2$ rapidly increases near $\dot{m} \sim \dot{m}_{\rm low}$,
the value of $\dot{m}_{\rm low}$ little depends on 
whether we define it by 
e.g., $r_2=10 r_{\rm g}$ or $r_2=20 r_{\rm g}$.

\subsubsection{Spectrum of gap emission}
\label{sec:stellar_SED}
\begin{figure}
  \includegraphics[angle=0,scale=0.40]{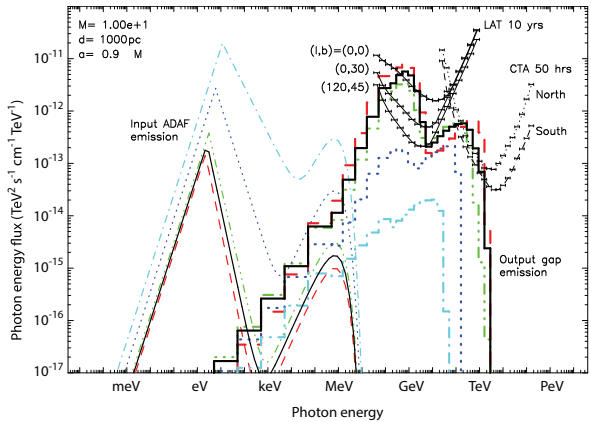}
\caption{
SED of the gap emission 
for a stellar-mass black hole with $M=10 M_\odot$ and $a_\ast=0.9$,
for five discrete dimensionless accretion rates, $\dot{m}$'s
at 1~kpc.
The thin curves denote the input ADAF spectra,
while the thick lines do the output gap spectra.
Each color corresponds to the same cases of $\dot{m}$'s
as in figures~\ref{fig:crtd_1e1}--\ref{fig:Ell_1e1}.
The thin solid curves (with horizontal bars)
denote the Fermi/{\it LAT} detection limits after 10 years observation,
while 
the thin dashed and dotted curves (with horizontal bars)
denote the CTA detection limits after a 50~hours observation.
Magnetic field strength is assumed to be the equipartition value
with the plasma accretion.
\label{fig:SEDa_1e1}
}
 \end{figure}

The predicted photon spectra are depicted in figure~\ref{fig:SEDa_1e1}
for the same set of $\dot{m}$ 
as in figures~\ref{fig:crtd_1e1} and \ref{fig:Ell_1e1}.
The thin curves on the left denote the input ADAF spectra,
while the thick lines on the right do the output spectra from the gap.
We find that the emitted flux increases with decreasing $\dot{m}$,
because the potential drop in the gap increases with decreasing $\dot{m}$.
The spectral peak around GeV is due to the curvature emission,
while that around TeV is due to the IC scatterings.
Provided that the distance is within several kpc,
these HE and VHE fluxes appear above the Fermi/{\it LAT}
detection limits 
(three thin solid curves labeled with \lq\lq LAT 10 yrs''),
\footnote{https://www.slac.stanford.edu/exp/glast/groups/canda/\\
          lat\_Performance.htm}
and the CTA detection limits 
(dashed and dotted curves labeled with \lq\lq CTA 50 hrs'').
\footnote{https://portal.cta-observatory.org/CTA\_Observatory\\
          /performance/SieAssets/SitePages/Home.aspx}
Since BH transients spend bulk of their time in 
a (very weakly accreting) quiescent state
\citep[e.g.,][]{miller11,plotkin13},
it is possible for near-by BH transients 
to exhibit detectable BH emissions.
For example, in HE, the flraring photons will be detectable
if the duty cycle of the flaring activities
is not too small (e.g., $>0.1$).

\begin{figure}
  \includegraphics[angle=0,scale=0.40]{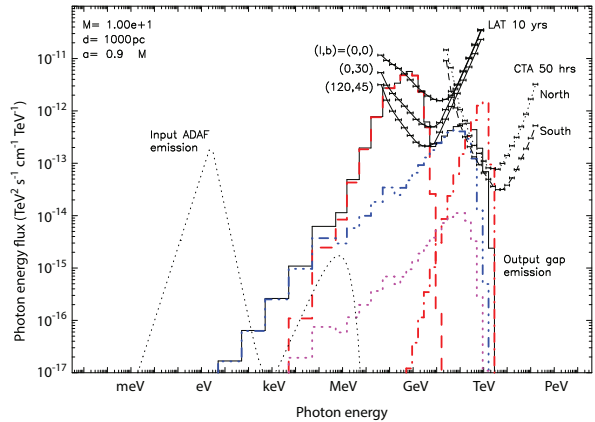}
\caption{
Similar figure as fig.~\ref{fig:SEDa_1e1},
but only the case of $\dot{m}=7.49 \times 10^{-5}$ is depicted.
The thick and thin black curves correspond to the same ones
in figure~\ref{fig:SEDa_1e1}.
The red dashed and dash-dotted lines denote
the primary curvature and inverse-Compton components, respectively.
The blue dash-dot-dot-dotted and purple dotted ones 
do the secondary and tertiary emission 
via synchrotron and inverse-Compton processes
outside the gap.
\label{fig:SEDb_1e1}
}
 \end{figure}

We plot the individual emission components in figure~\ref{fig:SEDb_1e1},
picking up the case of $\dot{m}=7.49 \times 10^{-5}$
(i.e., the case of the black solid line in fig.~\ref{fig:SEDa_1e1}).
The red dashed line shows the primary curvature component,
while the red dash-dotted line the primary IC component.
The former component is not absorbed and appear as the 
spectral peak at several GeV when $\dot{m} \sim \dot{m}_{\rm low}$
(i.e., when the gap outer boundary is located at $r_2 \gg r_{\rm g}$).
The latter component is heavily absorbed by the ADAF near-IR
photons to be reprocessed as the secondary component 
(blue dash-dot-dot-dotted line).
In this secondary component, IC emission dominates above GeV
and the synchrotron component dominates only below this energy.
The secondary component above 500 GeV is absorbed again
to be reprocess as the tertiary component
(purple dotted).

To grasp the strength of the absorption
taking place inside and outside the gap,
we examine the optical depth for photon-photon collisions.
For presentation purpose, we compute a representative
optical depth for test photons emitted outwards 
at the gap inner boundary, $r=r_1$.
With vanishing angular momenta,
photons propagate in the ($r$,$\theta$) surface 
(with constant $\theta$).
Thus, the invariant distance, $ds$,
of a photon path after propagating $dr$ and $d\varphi$
in $r$ and $\varphi$ coordinates becomes
\begin{equation}
  ds^2
  = g_{rr}dr^2+g_{\theta\theta}d\theta^2
  = \frac{\Sigma}{\Delta} dr^2
    \left[ 1 + O\left(\frac{\Delta\sin^2\theta}{r_{\rm g}^2}
                \right) 
    \right] .
\end{equation}
Thus, near the pole, $\vert\theta\vert \ll 1$, and
near the horizon, $\Delta \ll r_{\rm g}^2$,
we can put $ds \approx \sqrt{\Sigma/\Delta} dr$.
Integrating the local absorption probability over the photon ray,
we can compute the optical depth by
\begin{equation}
  \tau(\nu)
  = \int_{s_1}^\infty
      \frac{d\tau}{ds} ds
  = \int_{r_1}^\infty
      \sqrt{\frac{\Sigma}{\Delta}} \frac{d\tau}{ds} dr,
  \label{eq:tau}
\end{equation}
where $ds$ coincides with the distance of radial interval, $dr$,
measured by ZAMO.
The radial gradient of the optical depth becomes
\begin{equation}
  \frac{d\tau}{ds}
  = \frac{1}{c}
    \int_0^1 d\mu'
    \int_{\nu{}_{\rm th}'}^\infty 
      \frac{d\sigma_{\gamma\gamma}(\nu',\nu_{\rm s}',\mu')}{d\mu'}
      \frac{dF_{\rm s}'}{d\nu_{\rm s}'}
      d\nu_{\rm s}',
  \label{eq:tau2}
\end{equation}
where the primes denote quantities evaluated by ZAMO; 
$\mu'$ is the cosine of the collision angle of two photons,
$\nu_{\rm s}'$ the soft photon energy, and
$F_{\rm s}'$ the soft photon number flux.
We employ ZAMO here,
because the photons are assumed to be emitted with
vanishing angular momenta in this paper.
The threshold energy is defined by
$\nu_{\rm th}'=[2/(1-\mu')](m_{\rm e}c)^2/h^2\nu'$,
where $h$ means the Planck constant.
The photon frequency $\nu$ (at infinity) is de-redshifted 
to the ZAMO's value, $\nu'$ at each altitude, $r$.
Specifically, 
the local photon energy, $h\nu'$, is related to $h\nu$, by
$h\nu'= \dot{t} (h\nu+m\cdot d\varphi/dt)$,
where $m$ denotes the photon angular momentum 
and $d\varphi/dt$ the local observer's angular frequency 
in $dt$-basis at $r$.
However, we here have $m=0$, which gives the ZAMO's angular
frequency, $d\varphi/dt=-g_{t\varphi}/g_{\varphi\varphi}$.
The quantity $\dot{t}$ is given by the definition 
of the proper time, 
$\dot{t}^2
 [g_{tt}+2g_{t\varphi}d\varphi/dt
 +g_{\varphi\varphi}(d\varphi/dt)^2]= -1$,
which reduces to $\dot{t}=\sqrt{g_{\varphi\varphi}}/\rho_{\rm w}$
for a ZAMO.
Thus, we obtain 
$\nu'= \nu \sqrt{g_{\varphi\varphi}}/\rho_{\rm w}$
as the redshift relation between us and ZAMO.
Note that the integral $\int_{r_{\rm H}}^r \Delta^{-1/2}dr$
in equation~(\ref{eq:tau}) is finite for a finite $r$,
and that $dF_{\rm s}'/d\nu_{\rm s}'$,
and hence $d\tau/ds$ vanishes at large distances.

\begin{figure}
  \includegraphics[angle=0,scale=0.40]{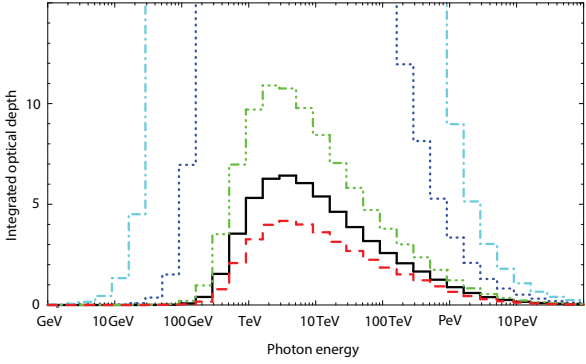}
\caption{
Optical depth, $\tau$, for photon-photon absorption
as a function of the photon energy, $h\nu$, at infinity
for five $\dot{m}$'s.
A test photon is supposed to be emitted 
outwards with zero angular momentum 
from the gap inner boundary to infinity. 
The photon energy $h\nu$ is de-redshited at each 
altitude, $r$, 
and the absorption probability gradient $d\tau/ds$ 
is integrated from the inner boundary to $60 r_{\rm g}$,
which is large enough.
\label{fig:f_tau_1e1}
}
 \end{figure}

In figure~\ref{fig:f_tau_1e1},
we present $\tau(\nu)$ for five discrete accretion rates:
The five lines correspond to the same cases of $\dot{m}$'s
as in figures~\ref{fig:crtd_1e1},
\ref{fig:Ell_1e1}, and \ref{fig:SEDa_1e1}.
It follows that the photon-photon absorption optical depth
exceeds unity above 16~GeV, 90~GeV, 0.3~TeV, 0.5~TeV, and 0.9~TeV
for $\dot{m}=1.00 \times 10^{-3}$,
$3.16 \times 10^{-4}$, $1.00 \times 10^{-4}$, $7.49 \times 10^{-5}$, 
and $5.62 \times 10^{-5}$, respectively.
The optical depth peaks at several TeV,
because the ADAF photon spectrum peaks in near-IR wavelengths
for stellar-mass BHs.

It should be stressed that the actual photons are absorbed 
by smaller optical depths than this figure,
because the individual photons are emitted at different 
positions whose altitudes are always higher than the inner boundary.

\subsubsection{Curvature versus Inverse-Compton processes}
\label{sec:stellar_processes}
\begin{figure}
  \includegraphics[angle=0,scale=0.40]{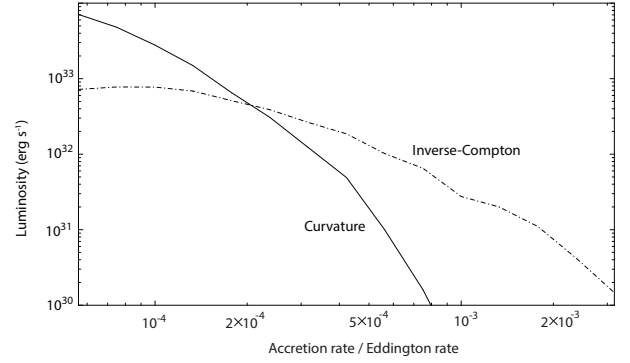}
\caption{
Gap luminosity as a function of the dimensionless accretion rate
for a stellar-mass black hole with $M=10 M_\odot$ and $a_\ast=0.9$.
The solid and dash-dotted curves denote the luminosity 
of the curvature and inverse-Compton processes, respectively.
\label{fig:luminosity_1e1}
}
 \end{figure}

It is worth examining the relative importance of the curvature
and IC processes.
In figure~\ref{fig:luminosity_1e1}, we plot the luminosity of the
outward curvature-emitted photons as the solid curve,
and that of the IC-emitted photons as the dash-dotted one,
as a function of $\dot{m}$.
The outward curvature photons are mostly emitted inside the gap,
while the IC photons are emitted both inside and outside the gap.
Figure~\ref{fig:luminosity_1e1} shows that 
the curvature luminosity exceeds the IC one 
when $\dot{m}<2 \times 10^{-4}$, or equivalently
when the gap extends enough.
This is because the curvature power is proportional to $\gamma^4$,
while the IC power approximately to $\gamma^0 \sim \gamma^2$, 
depending on whether the collisions take place 
mainly in the extreme Klein-Nishina or the Thomson regime.
The IC power also depends on the specific intensity of 
the soft photon field; thus, its dependence on $\gamma$ 
is more complicated than the curvature process.

\subsubsection{Luminosity versus created current density}
\label{sec:stellar_current}
\begin{figure}
  \includegraphics[angle=0,scale=0.40]{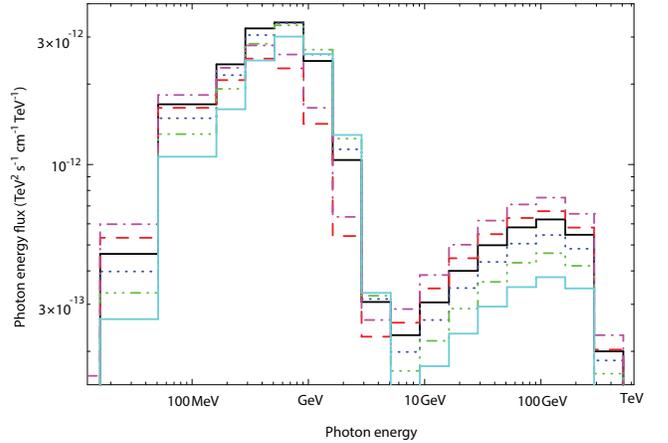}
\caption{
SED of the gap emission along the magnetic axis, $\theta=0$,
at the distance of 1~kpc,
from a BH with $M=10 M_\odot$ and $a=0.9 r_{\rm g}$.
The cyan solid, green dash-dot-dot-dotted,
blue dotted, black solid, red dashed, and purple dash-dotted lines
correspond to the dimensionless created current densities,
$j=0.4$, $j=0.5$, $0.6$, $0.7$, $0.8$, and $0.9$, respectively.
The accretion rate is fixed at $\dot{m}=1.00 \times 10^{-4}$.
\label{fig:SED_j_1e1}
}
\end{figure}

Let us briefly examine the dependence of the spectrum on the
created current density, $j$.
In figure~\ref{fig:SED_j_1e1}, 
we plot the gap spectra at six discrete $j$'s.
It follows that the HE emission becomes most luminous when
$0.5 \le j \le 0.7$
and the VHE one does when $j \sim 0.9$.
In what follows, we thus adopt $j=0.7$ as a compromise
to optimize the HE and VHE fluxes.
Note that we restrict our argument for $\vert j \vert \le 1$,
because $\vert j \vert > 1$ would incur a sign reversal of the 
$\rho-\rhoGJ$ in equation~(\ref{eq:pois_3}),
thereby resulting in a sign reversal of $E_\parallel$
at the outer boundary, $r=r_2$,
which would violate the present assumption of stationarity.

\subsubsection{Luminosity versus BH spin}
\label{sec:stellar_spin}
\begin{figure}
  \includegraphics[angle=0,scale=0.40]{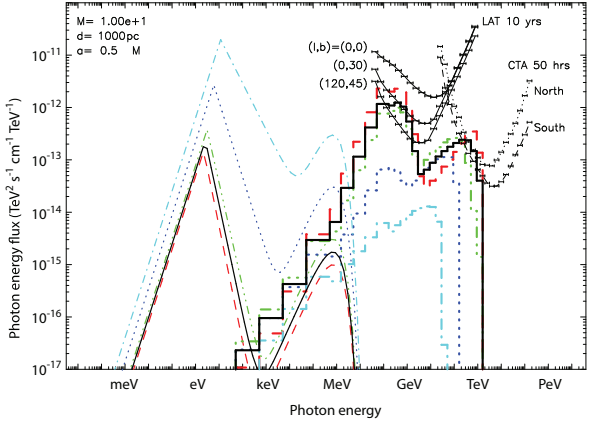}
\caption{
Similar figure as figure~\ref{fig:SEDa_1e1},
but the BH spin is chosen to be $a_\ast=0.5$,
while other parameters are the same,
including the accretion rate, $\dot{m}$.
\label{fig:SED_a500_1e1}
}
 \end{figure}

\begin{figure}
  \includegraphics[angle=0,scale=0.40]{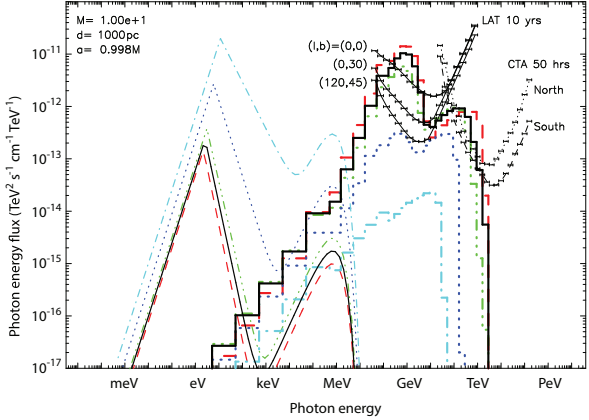}
\caption{
Similar figure as figures~\ref{fig:SEDa_1e1} and \ref{fig:SED_a500_1e1},
but the BH is maximally rotating, $a_\ast=0.998$.
Other parameters are the same as 
figures~\ref{fig:SEDa_1e1} and \ref{fig:SED_a500_1e1},
including the accretion rate, $\dot{m}$.
\label{fig:SED_a998_1e1}
}
 \end{figure}

We next briefly investigate how the spectrum depends on the
BH spin parameter, $a_\ast=a/r_{\rm g}$.
In figures~\ref{fig:SED_a500_1e1} and \ref{fig:SED_a998_1e1},
we present the gap spectra for $a_\ast=0.5$ and $0.998$
when $\dot{m}=1.00 \times 10^{-4}$.
It follows that the gap flux increases with increasing BH spin.
Thus, we adopt $a_\ast=0.9$ as the representative value
in this paper.

\subsubsection{Luminosity versus magnetic field rotation}
\label{sec:stellar_omegaF}
\begin{figure}
  \includegraphics[angle=0,scale=0.40]{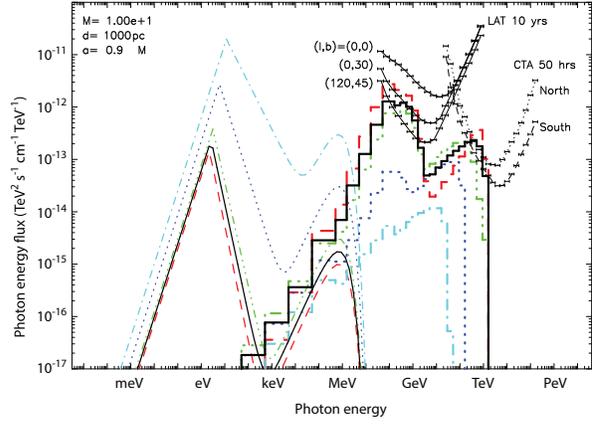}
\caption{
Similar figure as figures~\ref{fig:SEDa_1e1},
but the magnetosphere is more slowly rotating,
$\Omega_{\rm F}=0.20 \omega_{\rm H}$.
Other parameters are the same,
including the accretion rate, $\dot{m}$.
\label{fig:SEDa_omg02_1e1}
}
 \end{figure}

It is numerically suggested that the angular frequency, 
$\Omega_{\rm F}$, of an accreting BH magnetosphere 
decreases from $0.3 \omega_{\rm H}$ in the middle latitudes
to $-0.17 \omega_{\rm H}$ in the higher latitudes 
(i.e., near the pole) \citep{mckinney12}.
Similar tendency, from $0.4 \omega_{\rm H}$ to $-0.2 \omega_{\rm H}$ 
is, indeed, analytically suggested \citep{bes13}.
Thus, in this subsection, we examine a smaller $\Omega_{\rm F}$ case.
In figure~\ref{fig:SEDa_omg02_1e1},
we plot the SED for $\Omega_{\rm F}=0.2 \omega_{\rm H}$,
keeping other parameters unchanged from figure~\ref{fig:SEDa_1e1}.
It follows that the HE flux decreases to about 17~\% of the
$\Omega_{\rm F}=0.5 \omega_{\rm H}$ case,
while the VHE flux is roughly unchanged.
The VHE flux changes mildly,
because the IC process depends on the lepton Lorentz factor 
weakly, compared to the curvature process.
Note that the BZ power, which is proportional to
$\Omega_{\rm F}(\omega_{\rm H}-\Omega_{\rm F})$,
reduces to only 64~\% of the 
$\Omega_{\rm F}=0.5 \omega_{\rm H}$ case.
It means that the gap becomes less efficient
at smaller (in fact, also at greater) $\Omega_{\rm F}/\omega_{\rm H}$
than $0.5$.
Thus, we adopt $\Omega_{\rm F}=0.5\omega_{\rm H}$ 
as the representative value.

\subsubsection{Cascaded pairs outside gap}
\label{sec:stellar_cascaded}
\begin{figure}
  \includegraphics[angle=0,scale=0.40]{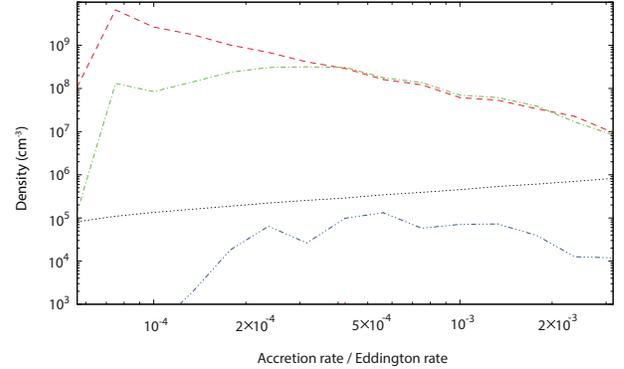}
\caption{
Density of the cascaded pairs outside the gap,
within radius $r_2 < r < 60r_{\rm g}$, 
is plotted as a function of $\dot{m}$ for $M=10 M_\odot$
and $a_\ast=0.9$.
The red dashed, green dash-dotted, and blue dash-dot-dot-dotted
curves denote the the densities
of the secondary, tertiary, and quaternary, 
pairs, respectively.
The black dotted curve shows the Goldreich-Julian charge density,
which is volume-averaged within the sphere of $r=60r_{\rm g}$.
\label{fig:pairs_1e1}
}
 \end{figure}

Let us investigate if a force-free magnetosphere is realized
outside the gap.
We compute the densities of the cascaded pairs
within the cutting radius $r_{\rm cut}=60r_{\rm g}$, 
which is well above the gap outer boundary, $r=r_2$.
In figure~\ref{fig:pairs_1e1},
we plot the secondary, tertiary, and quaternary pair densities
by the red dashed, green dash-dotted, and blue dash-dot-dot-dotted 
curves, respectively.
Here, the secondary pairs denote those cascaded from the primary
$\gamma$-rays, which are defined to be emitted by the primary
electrons or positrons that are accelerated in the gap.
The tertiary pairs denote those cascaded from the secondary
$\gamma$-rays, which are defined to be emitted 
by these secondary pairs. 

It follows that the density of the cascaded pairs 
exceeds the GJ one (black dotted line), 
as long as a stationary gap is formed.
Thus, the magnetosphere becomes force-free outside the gap,
as long as $\dot{m} > \dot{m}_{\rm low}$.
This conclusion does not depend on the choice of $r_{\rm cut}$.

In short, gap solutions exist if the dimensionless accretion rate
is in the range $\dot{m}_{\rm low} < \dot{m} < \dot{m}_{\rm up}$.
For stellar-mass BHs, the gap emission peaks at several GeV 
and its maximum  flux $> 10^{-12} \mbox{ ergs s}^{-1}$ 
is detectable with {\it Fermi}/LAT,
if the duty cycle of the flaring activity is not too small.
The gap luminosity increases with decreasing $\dot{m}$,
because the gap is dissipating a portion of the BH's 
spin-down luminosity.
This forms a striking contrast to accretion-powered systems,
whose luminosity will decrease with decreasing $\dot{m}$.
The cascaded pairs outside the gap have a greater density
than the Goldreich-Julian value;
thus, the magnetosphere becomes force-free 
in the downstream of the gap-generated flow
(i.e., outside the gap outer boundary).


\subsection{The case of intermediate mass BHs}
\label{sec:IMBH}
\begin{figure}
  \includegraphics[angle=0,scale=0.40]{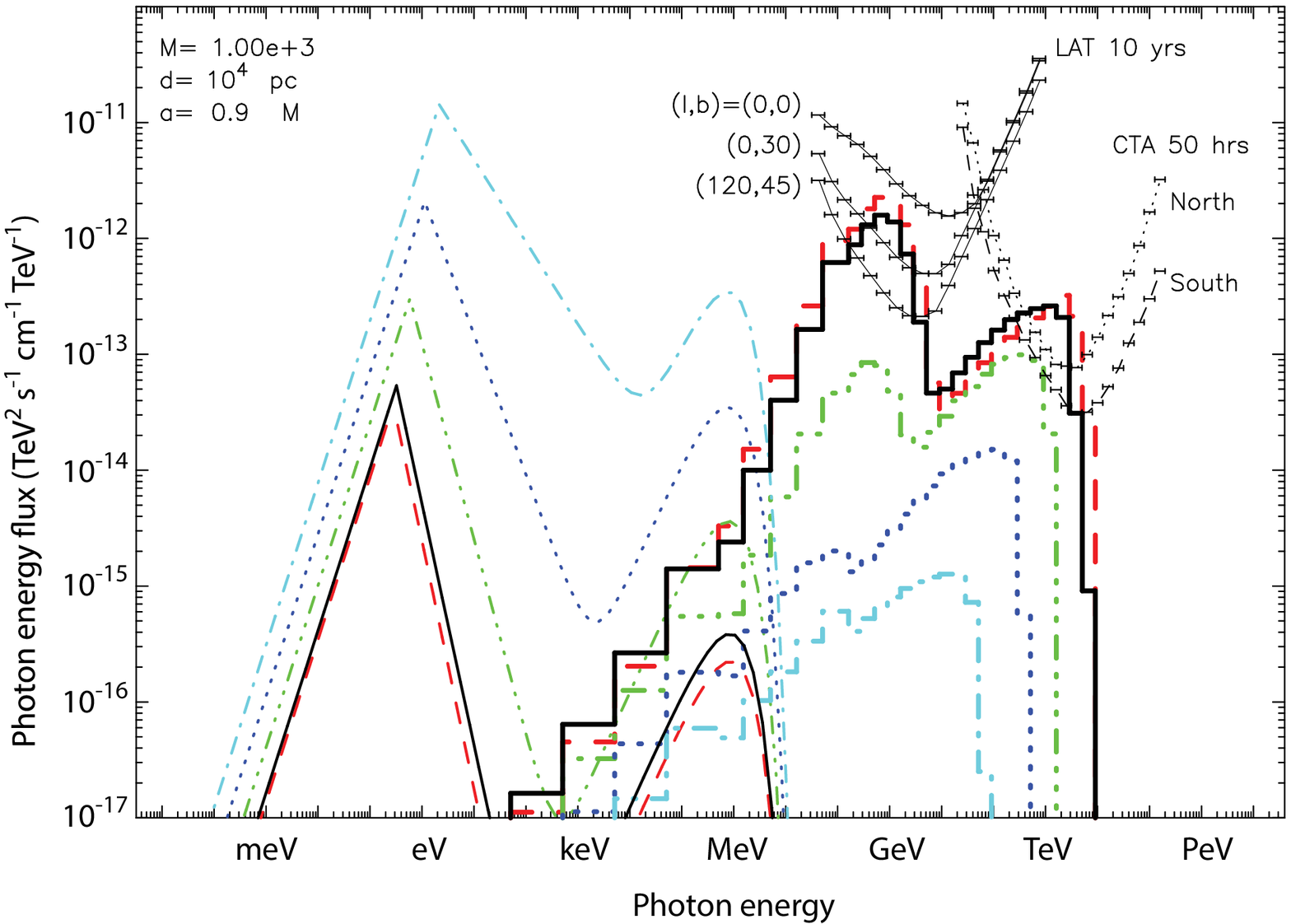}
  \includegraphics[angle=0,scale=0.40]{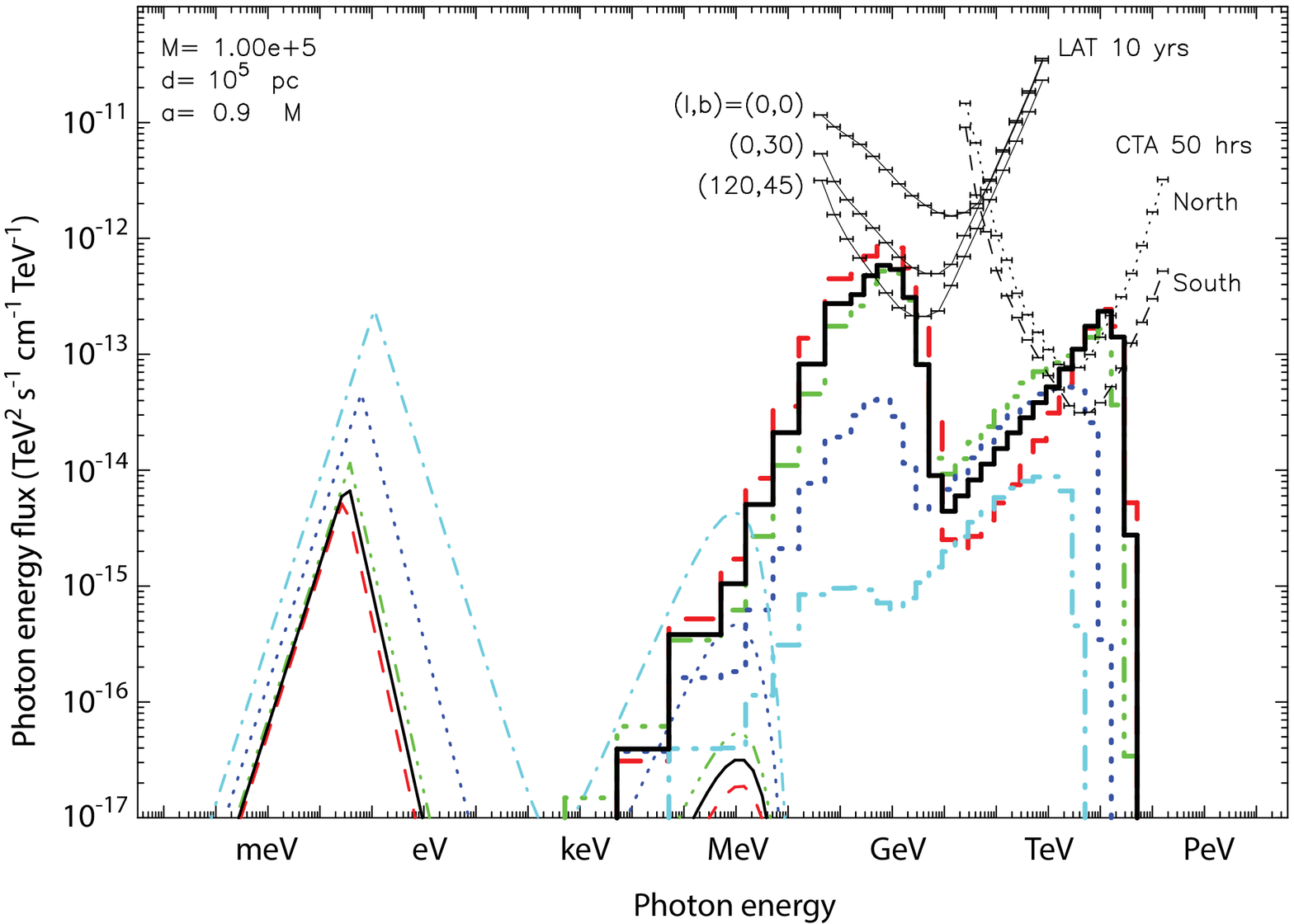}
\caption{
{\it Top panel:}
SED of the gap for a BH with
$M=10^3 M_\odot$, $a_\ast=0.9$, and $\Omega_{\rm F}=0.5\omega_{\rm H}$.
The distance is assumed to be 10~kpc.
The cyan, blue, green, black, and red
curves represent the SEDs for
$\dot{m}=1.00 \times 10^{-3}$,
$3.16 \times 10^{-4}$,
$1.00 \times 10^{-4}$,
$3.16 \times 10^{-5}$, and 
$2.37 \times 10^{-5}$, respectively.
{\it Bottom panel:}
SED for 
$M=10^5 M_\odot$, $a_\ast=0.9$, and $\Omega_{\rm F}=0.5\omega_{\rm H}$.
The distance is assumed to be 100~kpc.
The cyan, blue, green, black, and red
curves represent the SEDs for
$1.00 \times 10^{-4}$,
$\dot{m}=3.16 \times 10^{-5}$,
$1.00 \times 10^{-5}$,
$7.49 \times 10^{-6}$, and 
$5.62 \times 10^{-6}$, respectively.
\label{fig:SEDa_1e5}
}
 \end{figure}

Let us examine the gap emission from intermediate-mass BHs.
We put $M=10^3 M_\odot$ and calculate the gap emission
for $a_\ast=0.9$ and $\Omega_{\rm F}=0.5 \omega_{\rm H}$.
The predicted spectra become as 
the top panel in figure~\ref{fig:SEDa_1e5}
for a distance of $10$~kpc.
It shows that the $\gamma$-ray fluxes  are detectable
in both HE and VHE if an IMBH is located within our galaxy,
as long as the accretion rate is in the range, 
$2 \times 10^{-5} < \dot{m} < 4 \times 10^{-5}$.

Since the absolute luminosity of the gap increases with $M$,
we also compute the SEDs for $M=10^5 M_\odot$ 
(bottom panel of fig.~\ref{fig:SEDa_1e5}),
assuming $d=100$~kpc.
It shows that such a heavy BH is detectable in HE or VHE
only when the distance is comparable to or less than $100$~kpc.

\begin{figure}
  \includegraphics[angle=0,scale=0.40]{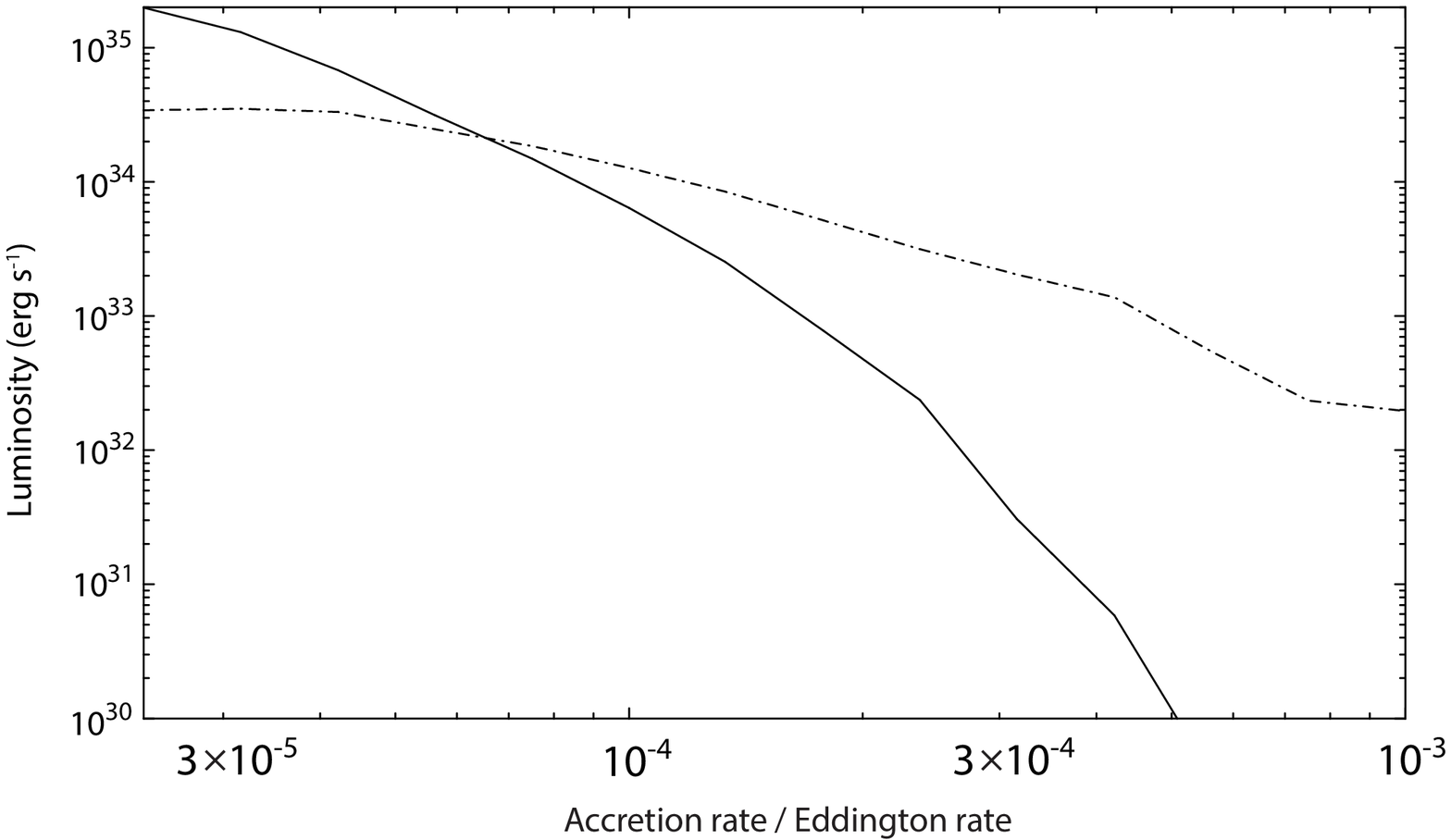}
  \includegraphics[angle=0,scale=0.40]{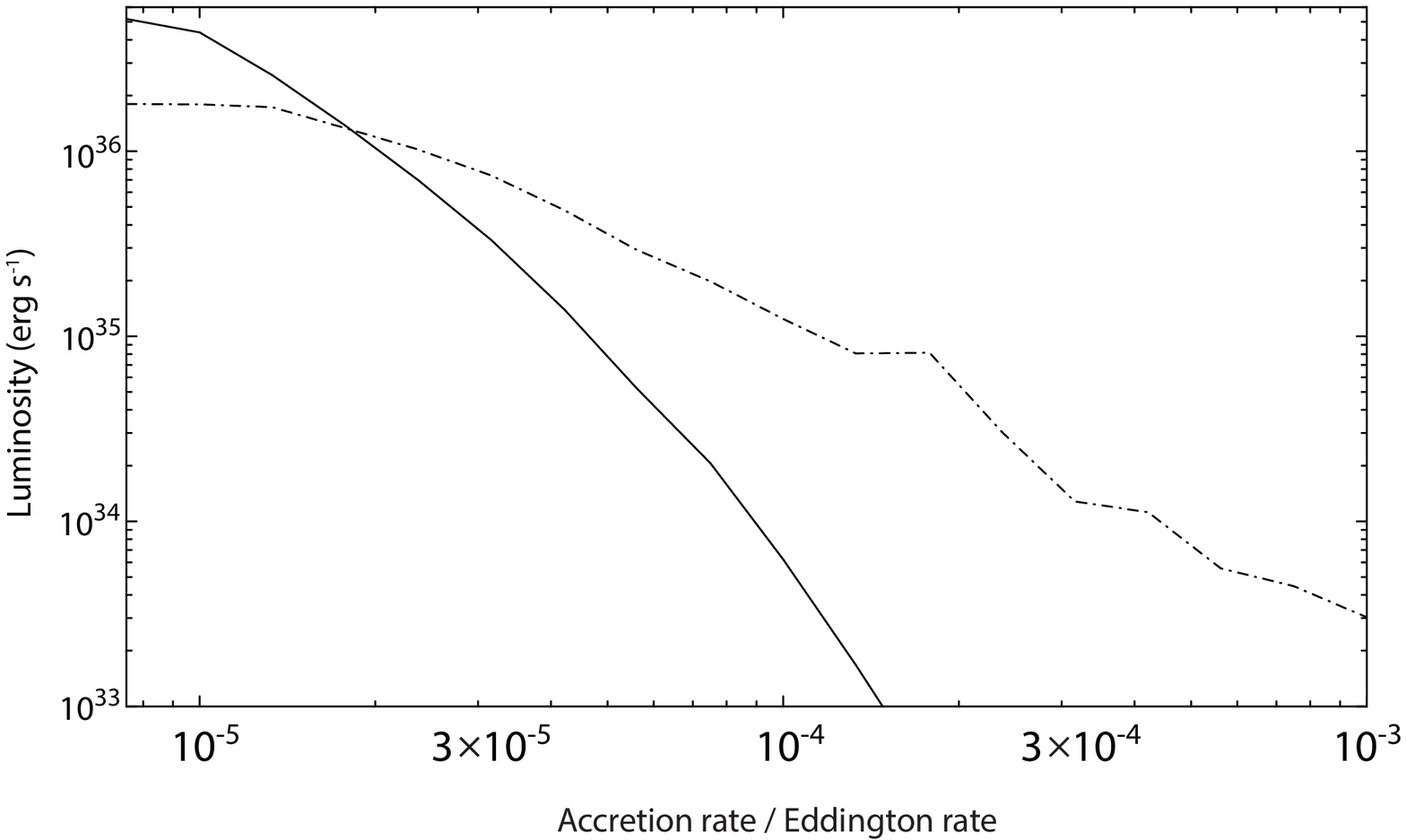}
\caption{
{\it Top panel:}
Gap luminosity as a function of the dimensionless accretion rate
for a BH with 
$M=10^3 M_\odot$, $a_\ast=0.9$, and $\Omega_{\rm F}=0.5\omega_{\rm H}$.
The solid and dash-dotted curves denote the luminosity 
of the curvature and inverse-Compton processes, respectively.
{\it Bottom panel:}
Similar figure as the top panel but for $M=10^5 M_\odot$.
\label{fig:luminosity_1e5}
}
 \end{figure}

The curvature and IC luminosities are plotted as a function of
$\dot{m}$ in figure~\ref{fig:luminosity_1e5}.
It shows that the curvature process works stronger than the IC 
when $\dot{m}< 6   \times 10^{-5}$ for $M=10^3 M_\odot$, and 
when $\dot{m}< 1.5 \times 10^{-5}$ for $M=10^5 M_\odot$.

\begin{figure}
  \includegraphics[angle=0,scale=0.40]{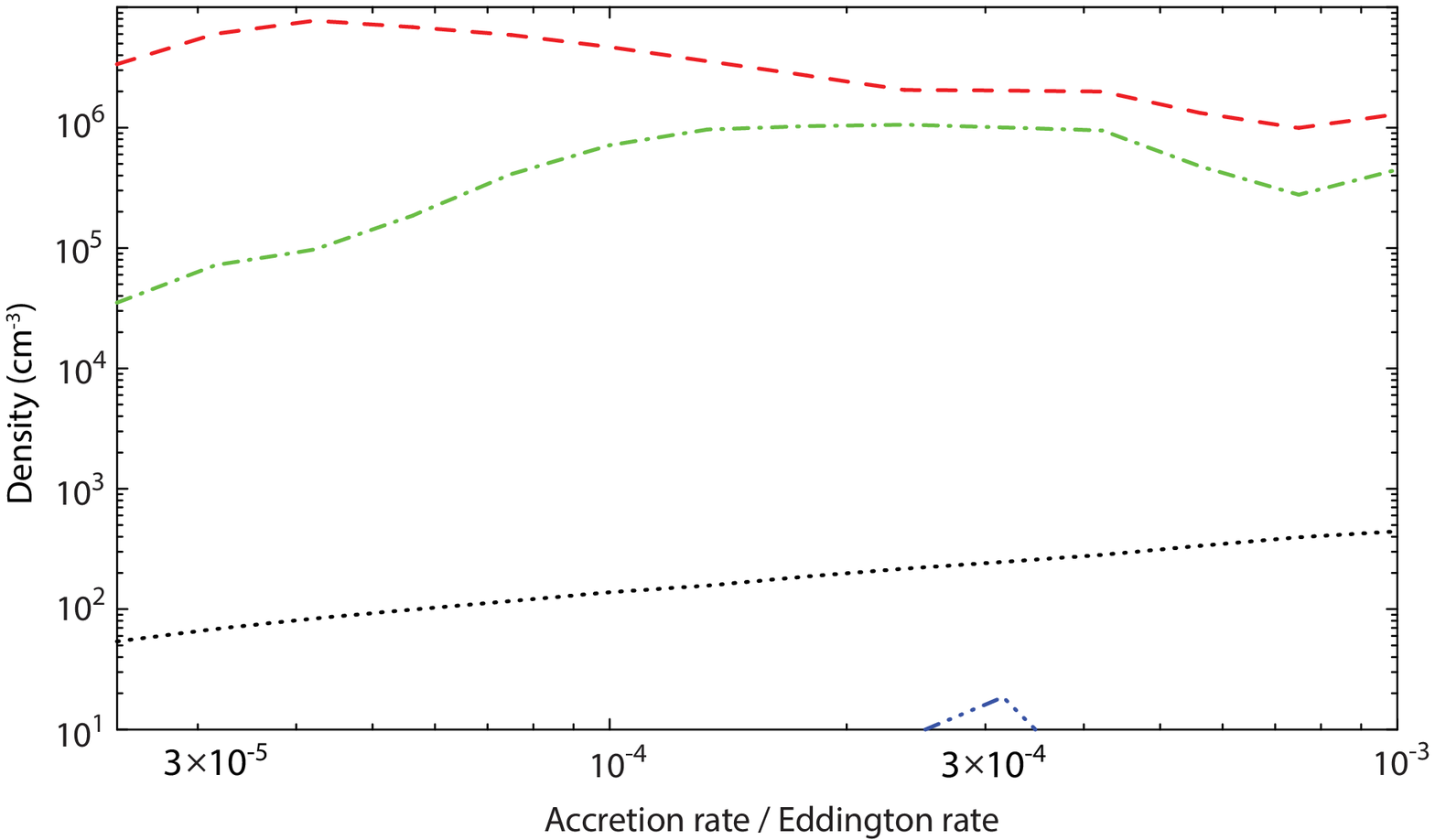}
  \includegraphics[angle=0,scale=0.40]{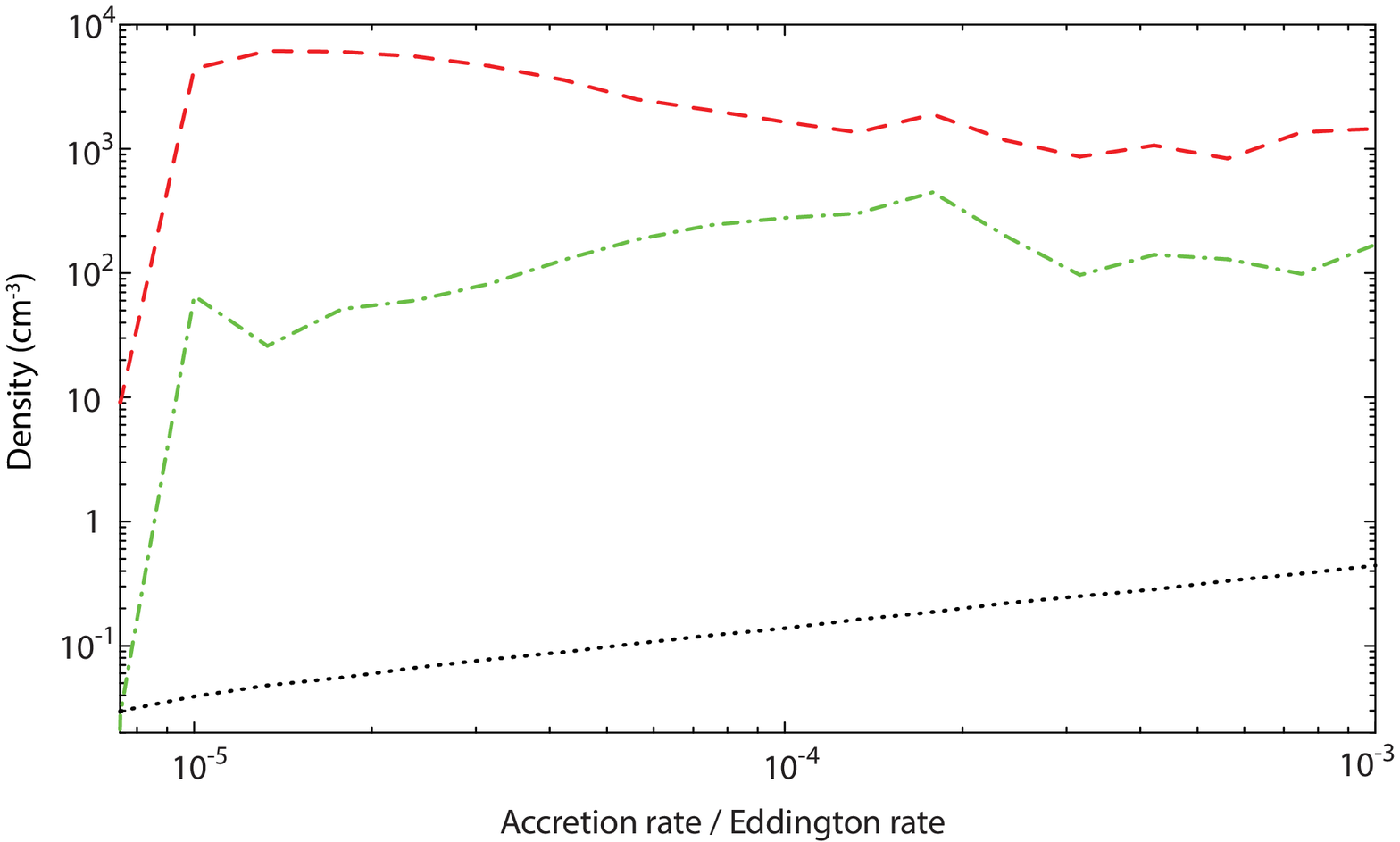}
\caption{
Similar figure as figure~\ref{fig:pairs_1e1},
but for a BH with $M=10^3 M_\odot$ (top panel)
and $M=10^5 M_\odot$ (bottom panel); 
$a_\ast=0.9$ and $\Omega_{\rm F}=0.5\omega_{\rm H}$ are unchanged.
The dashed and dash-dotted curves denote 
the densities, $N_\pm$, of the secondary and tertiary pairs,
while the dotted one the GJ value, $N_{\rm GJ}$.
\label{fig:pairs_1e5}
}
 \end{figure}

We examine if a force-free magnetosphere is sustained.
Figure~\ref{fig:pairs_1e5} shows the densities of the pairs
cascaded between $r_2$ and $60r_{\rm g}$.
It follows that the density of the cascaded pairs 
exceeds the GJ one for both $M=10^3 M_\odot$ and $M=10^5 M_\odot$.

\subsection{The case of super-massive BHs}
\label{sec:SMBH}
Next, let us examine super-massive BHs.
We present the acceleration electric field 
in \S~\ref{sec:SMBH_Ell},
gap spectrum 
in \S~\ref{sec:SMBH_SED},
and cascaded pair densities 
in \S~\ref{sec:SMBH_cascaded}.
We briefly examine the dependence on the
soft photon density
in \S~\ref{sec:SMBH_dep_RIAF}.

\subsubsection{Acceleration electric field}
\label{sec:SMBH_Ell}
\begin{figure}
  \includegraphics[angle=0,scale=0.70]{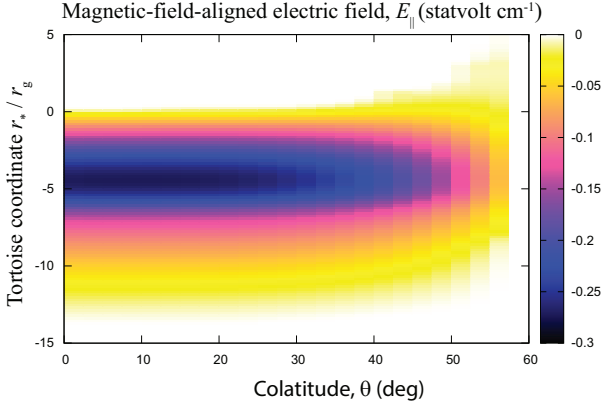}
\caption{
Similar figure as figure~\ref{fig:e2d_th_1e1},
but for $M=10^9 M_\odot$.
The dimensionless accretion rate is chosen to be
$\dot{m}=1.00 \times 10^{-6}$.
\label{fig:e2d_1e9}
}
 \end{figure}

\begin{figure}
  \includegraphics[angle=0,scale=0.40]{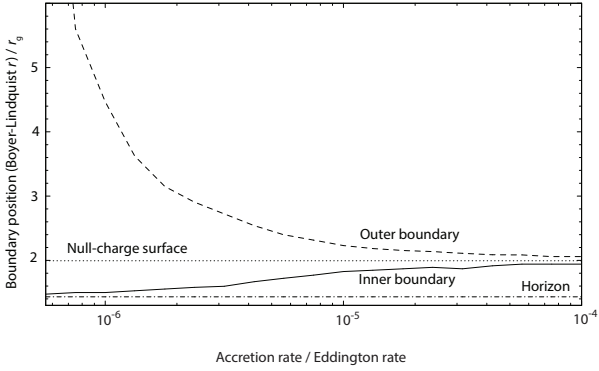}
\caption{
Similar figure as figure~\ref{fig:width_1e1},
but for a super-massive black hole with $M=10^9 M_\odot$
and $a_\ast=0.9$.
\label{fig:width_1e9}
}
\end{figure}

Choosing a typical mass of $M=10^9 M_\odot$,
we plot $E_\parallel(\eta_\ast,\theta)$ distribution
in figure~\ref{fig:e2d_1e9},
whose ordinate is the dimensionless tortoise coordinate, $\eta_\ast$.
Comparing with figure~\ref{fig:e2d_1e1},
we find that the essential behavior of $E_\parallel$
is unchanged from the case of $M=10 M_\odot$.
In Boyer-Lindquist radial coordinate,
the gap outer and inner boundaries distribute
as a function of $\dot{m}$ as depicted in figure~\ref{fig:width_1e9}.
The gap outer boundary is located at
$r_2=5.60 r_{\rm g}$ and $r_2=10.32 r_{\rm g}$
at $\dot{m} = 7.49 \times 10^{-7}$ and $5.62 \times 10^{-7}$,
respectively; 
thus, we obtain $\dot{m}_{\rm low} = 5.75 \times 10^{-7}$
for $M=10^9 M_\odot$, $a=0.90 r_{\rm g}$,
$\Omega_{\rm F}=0.50 \omega_{\rm H}$, and $j=0.7$. 
Figure~\ref{fig:width_1e9} shows 
that the gap longitudinal width becomes comparable or
greater than $r_{\rm g}$ when the accretion rate 
reduces to  $\dot{m} < 2 \times 10^{-6}$,
whereas it is realized when $\dot{m} < 2 \times 10^{-4}$
for a stellar-mass case (fig.~\ref{fig:width_1e1}).
It also follows that the gap outer boundary shifts
outwards with decreasing $\dot{m}$,
in the same manner as the $M=10 M_\odot$ case.

\subsubsection{Spectrum of gap emission}
\label{sec:SMBH_SED}
\begin{figure}
  \includegraphics[angle=0,scale=0.40]{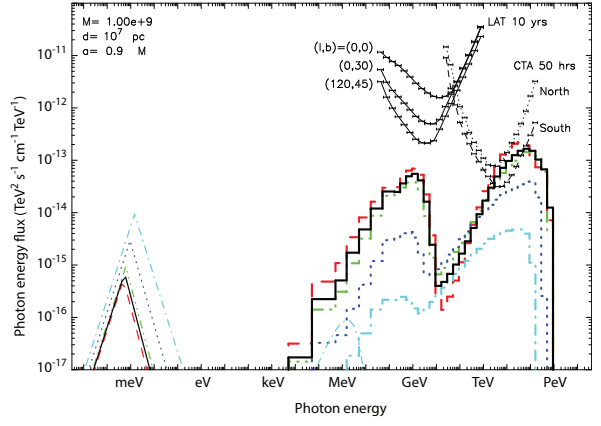}
\caption{
SED of the gap for a super-massive BH with
$M=10^9 M_\odot$ and $a_\ast=0.9$ at 10~Mpc.
The cyan, blue, green, black, and red
curves represent the SEDs for
$\dot{m}=1.00 \times 10^{-5}$,
$3.16 \times 10^{-6}$,
$1.00 \times 10^{-6}$,
$7.49 \times 10^{-7}$, and 
$5.62 \times 10^{-7}$, respectively.
\label{fig:SEDa_1e9}
}
 \end{figure}

Figure~\ref{fig:SEDa_1e9} shows the gap spectra
as the thick lines for five discrete $\dot{m}$'s, 
assuming a luminosity distance of $10$~Mpc.
When the accretion rate is in the range
$5.6 \times 10^{-7} < \dot{m} < 10^{-6}$,
we find that the gap emission will be marginally detectable 
with CTA,
if the source is located in the southern sky.

\begin{figure}
  \includegraphics[angle=0,scale=0.40]{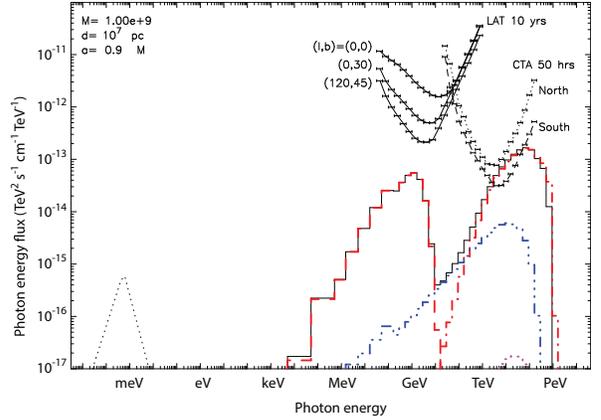}
\caption{
Similar figure as fig.~\ref{fig:SEDa_1e9},
but only the case of $\dot{m}=7.49 \times 10^{-7}$ is depicted.
The thick and thin black curves correspond to the same ones
in figure~\ref{fig:SEDa_1e9}.
The red dashed and dash-dotted lines denote
the primary curvature and inverse-Compton components, respectively.
The blue dash-dot-dot-dotted line
do the secondary emission 
via synchrotron and inverse-Compton processes
outside the gap.
\label{fig:SEDb_1e9}
}
 \end{figure}

The emission components are depicted in figure~\ref{fig:SEDb_1e9}
for $\dot{m}=7.49 \times 10^{-7}$.
Since $\dot{m}$ is much smaller than the stellar-mass cases,
the absorption optical depth decreases accordingly;
as a result,
most of the primary IC component (red dash-dotted line) cascades
only to the secondary generation pairs,
whose emission is represented by 
the blue dash-dot-dot-dotted line.

\begin{figure}
  \includegraphics[angle=0,scale=0.40]{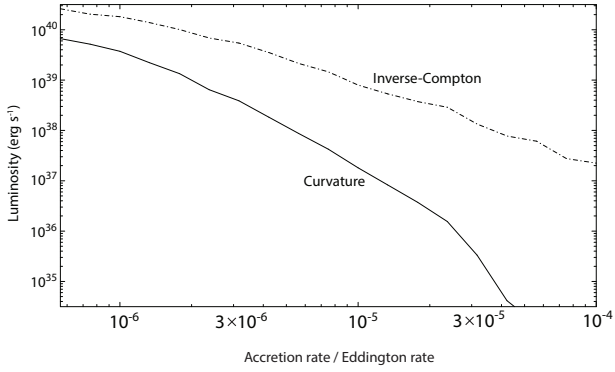}
\caption{
Gap luminosity as a function of the dimensionless accretion rate
for a super-massive black hole with $M=10^9 M_\odot$ and $a_\ast=0.9$.
The solid and dash-dotted curves denote the luminosity 
of the curvature and inverse-Compton processes, respectively.
\label{fig:luminosity_1e9}
}
 \end{figure}

The curvature and IC luminosities are plotted as a function of
$\dot{m}$ in figure~\ref{fig:luminosity_1e9}.
It is clear that the IC process dominates the curvature one
in the entire range of $\dot{m}$ for super-massive BHs.

\subsubsection{Cascaded pairs outside gap}
\label{sec:SMBH_cascaded}
\begin{figure}
  \includegraphics[angle=0,scale=0.40]{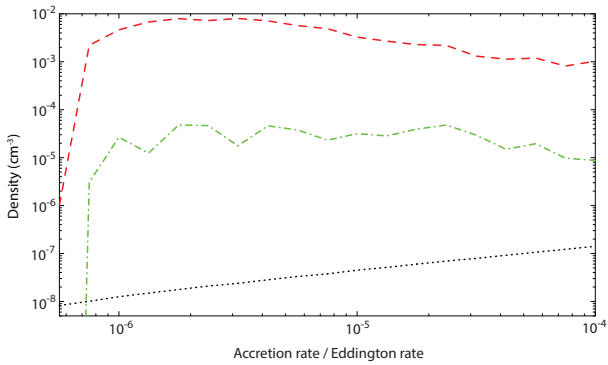}
\caption{
Similar figure as figure~\ref{fig:pairs_1e1},
but for a super-massive BH case, $M=10^9 M_\odot$.
The dashed and dash-dotted curves denote 
the densities, $N_\pm$, of the secondary and tertiary pairs,
while the dotted one the GJ value, $N_{\rm GJ}$.
\label{fig:pairs_1e9}
}
 \end{figure}

The created pair densities between $r_2$ and $60 r_{\rm g}$ 
are depicted as a function of $\dot{m}$ 
in figure~\ref{fig:pairs_1e9}.
In the same way as stellar-mass BHs (\S~\ref{sec:stellar})
and intermediate-mass BHs (\S\ref{sec:IMBH}),
the magnetosphere becomes force-free outside the gap,
as long as the gap solution exists.

Figures~\ref{fig:pairs_1e1}, \ref{fig:pairs_1e5}
and \ref{fig:pairs_1e9} show that
the BH magnetosphere becomes force-free for
$M=10 M_\odot$, $10^3 M_\odot$, $10^5 M_\odot$, and $10^9 M_\odot$.
Indeed, a BH magnetosphere becomes
force-free irrespective of the BH mass,
as long as a stationary gap is sustained in it by 
$\dot{m}>\dot{m}_{\rm low}(M)$.

\subsubsection{Dependence on RIAF soft photon field density}
\label{sec:SMBH_dep_RIAF}
\begin{figure}
  \includegraphics[angle=0,scale=0.40]{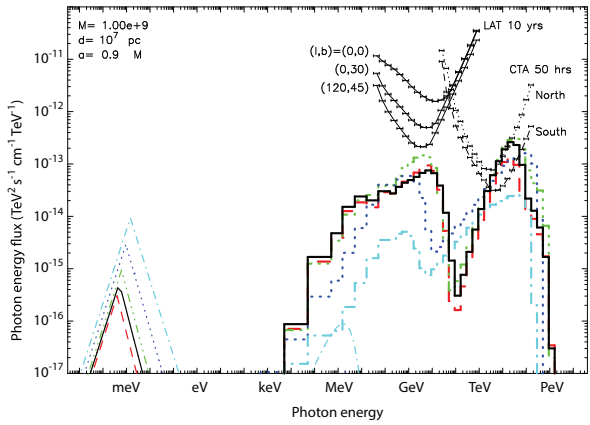}
\caption{
Similar figure as figure~\ref{fig:SEDa_1e9},
but the ADAF photon field is artificially reduced
to one-fourth by adopting $R_{\rm min}=12 r_{\rm g}$,
instead of $R_{\rm min}=6 r_{\rm g}$.
\label{fig:SEDa_12rg_1e9}
}
 \end{figure}

Let us quickly examine how the gap luminosity 
depends on the ADAF photon field density,
fixing $\dot{m}$ so that the magnetic field near the horizon
may not be changed.
As a test model, we artificially reduce the
photon density near the horizon to one-fourth of its original value
by doubling $R_{\rm min}$.
In figure~\ref{fig:SEDa_12rg_1e9},
we plot the gap spectra for $R_{\rm min}=12 r_{\rm g}$.
We find that the gap emission increases because of the diminished
soft photon density, as expected.
To predict the gap spectrum of SMBHs further precisely,
we must constrain the specific intensity of the RIAF photon field 
near the horizon such as by numerical simulations.  

\subsection{Gap luminosity for various BH masses}
\label{sec:general}
Finally, we apply the method to various BH masses
from $10 M_\odot$ to $6.4 \times 10^9 M_\odot$,
The filled circles in figure~\ref{fig:mdot}
denote the solved $\dot{m}_{\rm low}(M)$
for 
$(a_\ast,\Omega_{\rm F}/\omega_{\rm H},R_{\rm min}/r_{\rm g})
 =(0.9,0.5,6)$.
It follows that $\dot{m}_{\rm low}(M)$ always lies below
the upper limit, $\dot{m}_{\rm up}$ (upper straight line).
Therefore, BH gaps exist for arbitrary BH masses.

We fit the filled circles in figure~\ref{fig:mdot}
with a straight line by least square method to obtain
$\lg \dot{m}_{\rm low}=-3.6 -0.26 \lg(M/M_\odot)$.
The luminosity, $L_{\rm gap}$, of a BH gap, 
could be estimated by substituting
this $\dot{m} \approx \dot{m}_{\rm low}$ into 
$ L_{\rm gap} \approx 0.3 L_{\rm BZ} 
 \approx 1.5 \times 10^{38} \dot{m} M_1 \mbox{ergs s}^{-1}$,
where equation~(\ref{eq:L_BZ_2}) is used.
The factor $0.3$ comes from the fact that the gap 
luminosity can attain at most $\sim 30$~\% of $L_{\rm BZ}$,
because $E_\parallel$ is partly screened by the created pairs
within the gap, and because a stationary gap solution can possess
only a sub-GJ current density.
It is also noteworthy that the BH-gap solution at such a low $\dot{m}$ 
corresponds to the case of the middle-aged pulsars
whose outer-gap luminosity attains at most 30~\% 
of the spin-down luminosity from the same reasons \citep{hiro13}.

If we artificially change $R_{\rm min}$, 
we can consider the impact of the ADAF photon field density,
without changing $B$ at the same $\dot{m}$.
In figure~\ref{fig:mdot},
we plot the $\dot{m}_{\rm cr}(M)$ for 
$(a_\ast,\Omega_{\rm F}/\omega_{\rm H},R_{\rm min}/r_{\rm g})
 =(0.9,0.5,12)$ as the open squares.
It follows that the reduction of the soft photon field 
(by doubling $R_{\rm min}$) changes the gap solution
to some extent,
because it affects the pair production process in the gap.

In figure ~\ref{fig:mdot} we also plot a slower BH spin,
$(a_\ast,\Omega_{\rm F}/\omega_{\rm H},R_{\rm min}/r_{\rm g})
 =(0.5,0.5,6)$ as the open circles.
It follows that $\dot{m}_{\rm low}(M)$ has a weak dependence 
on $a_\ast$ due to the negative feedback effect \citep{hiro13}.
To further constrain $\dot{m}_{\rm low}(M)$,
we need to specify the specific intensity of the RIAF (e.g., ADAF) 
photon field near the horizon by numerical computation.
Such details are, however, out of scope of the present paper.
Therefore, the region $\dot{m}<\dot{m}(M)$ 
(in which stationary gaps do not exist)
is depicted in red {\it gradation} in figure~\ref{fig:mdot}.

\section{Discussion}
\label{sec:disc}
To sum up, 
we have solved stationary lepton accelerators (or gaps)
in the magnetospheres of rotating black holes with arbitrary masses.
By solving the set of an inhomogeneous part of the Maxwell equations,
lepton equations of motion, and the radiative transfer equation,
we demonstrate that an electric field arises along the magnetic field
line around the null charge surface
on which the frame-dragging angular frequency coincides with
the magnetic-field angular frequency.
In the gap, electrons and positrons are created via two-photon collisions
and accelerated in opposite directions by the acceleration electric field
into ultra-relativistic energies.
Such leptons emit copious $\gamma$-rays via curvature and 
inverse-Compton (IC) processes, 
leading to a pair-production cascade inside and outside the gap.
The gap longitudinal width is self-regulated so that a single electron
eventually cascade into a single pair within the gap,
and approximately coincides with the mean-free path 
(for an IC photon to materialize via two-photon collision)
divided by the number of IC photons emitted by a single electron.
As the accretion rate decreases,
the increased mean-free path results in an extended gap,
and hence an increased luminosity.
The gap luminosity maximizes when the gap width becomes
much greater than the horizon radius.
For stellar-mass BHs, we can expect that their curvature emission
are detectable with {\it Fermi}/LAT 
and the IC emission with {\it CTA},
when the BH binary is in a quiescent state,
provided that the distance is within several kpc.
For super-massive BHs, their IC emission may be marginally 
detectable with {\it CTA} for low luminosity AGN, 
if their distances are within a few tens of Mpc.

\subsection{Improvement form HP16}
\label{sec:disc_improvement}
In the present work, there are mainly two improvements from HP16, 
which formulated a BH gap model and applied it to 
the radio galaxy IC310.
First, in the present work, 
the Poisson equation~(\ref{eq:pois}) is solved
general-relativistically on the poloidal plane.
However, in HP16,
the left-hand side of equation~(\ref{eq:pois}) 
was approximated one-dimensionally in the Newtonian limit.
Second, in the present work, 
$\rho(r,\theta)$ (in the Poisson equation) 
is solved from the local pair production rate
in a consistent manner with the radiative transfer equation.
However, HP16 assumed that $\rho(r)$ changes linearly with $r$ 
(see their eq.~10).

\begin{figure}
  \includegraphics[angle=0,scale=0.40]{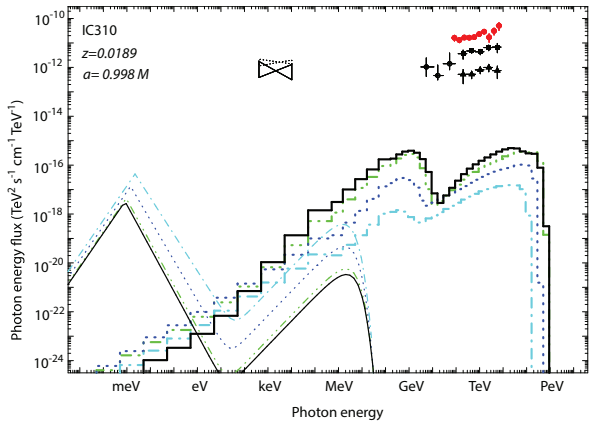}
\caption{
BH gap emission spectrum of IC310*. 
Extreme spin, $a_\ast=0.998$, is assumed 
to compare with figure~13 ($B=B_{\rm eq}$ case) of HP16.
The cyan, blue, green, and black curves correspond to
$\dot{m}=1.00 \times 10^{-5}$,
$3.16 \times 10^{-6}$,
$1.00 \times 10^{-6}$, and 
$7.49 \times 10^{-7}$, respectively.
The black triangles and squares
show the observed fluxes in the low and high states
\citep{alek11} respectively,
while the red circles do those in the flaring state \citep{alek14b}.
The observed VHE fluxes are corrected for the absorption
by the extragalactic background light.
The solid bowtie represents the X-ray fluxes
obtained by re-analyzing
the observational data taken with
{\it XMM-Newton} in 2003,
{\it Chandra} in 2004, and
{\it Swift} in 2007 \citep{alek14a}.
The dotted bowtie represents the re-analyzed data
of {\it Chandra} in 2005 \citep{alek14a}.
The open circles denote the {\it Fermi}/LAT spectrum.
\label{fig:SEDa_IC310}
}
 \end{figure}

Both above-mentioned simplifications suppress $E_\parallel$ 
to some extent, comparing the GR treatment.
To compare the difference,
we apply the present 2-D, GR method to IC310, assuming $B=B_{\rm eq}$.
Figure~\ref{fig:SEDa_IC310} show the resultant SED,
where the parameter set is basically the same as HP16;
namely, we adopt $a_\ast=0.998$ and $\Omega_{\rm F}=0.3 \omega_{\rm H}$.
Comparing with figure~13 of HP16, 
which was obtained for $B=B_{\rm eq}$,
we find that the 1-D Newtonian approximation in HP16
underestimated the gap luminosity about $1.5$ times from the
present 2-D GR treatment.

\subsection{Rigidity of rotating magnetic field lines}
\label{sec:disc_gididity}
We have assumed that 
$\Omega_{\rm F} \equiv F_{tr}/F_{r\varphi}= F_{t\theta}/F_{\theta\varphi}$ 
is constant along each magnetic field line.
However, $\Omega_{\rm F}$ is not conserved 
when the electric potential drop along the magnetic field line becomes a
non-negligible fraction of the electro-motive force exerted across the
horizon.
Nevertheless, for all the cases in the present paper,
the gap luminosity is less than two percent of the Blandford-Znajek power. 
Noting that the particle energy is predominantly converted into radiation 
within the gap, we can conclude that the deviation of $\Omega_{\rm F}$ 
from a constant value (e.g., $0.5 \omega_{\rm H}$) 
is at most a few percent.
Thus, the assumption of a constant $\Omega_{\rm F}$ along
each magnetic field line is mostly justified.

\subsection{Gap position versus the separation surface}
\label{sec:disc_separation}
It is worth comparing the gap position with the separation surface
where both inflows and outflows start without $E_\parallel$.
Since the plasma particles will not be
accelerated either outward or inward at this surface, 
the sum of the gravitational, Lorentz and centrifugal forces
will vanish along the magnetic field lines there.
One of the convenient ways to find such a surface is to put 
both $u_{\rm p}=0$ and $u_{\rm p}{}'=0$ in the poloidal wind equation 
of MHD \citep{came86a,came86b}, 
where $u_{\rm p}$ refers to the poloidal velocity of the fluid.
Note that this MHD argument is valid in the magnetically dominated limit,
which is close to (but not equal to) the force-free limit.

In a stationary and axisymmetric BH magnetosphere, 
the separation surface is determined by the condition 
$k_0'=0$ in MHD
\citep[see end of \S~\ref{sec:geometry};][]{taka90}, 
where the prime denotes the derivative along the poloidal 
magnetic field line. 
For a radial poloidal field lines, 
$k_0$ attains its maximum value
$0.651$, $0.478$, $0.372$, and $0.310$ 
at $r=8.34 r_{\rm g}$, $5.21 r_{\rm g}$, $4.02 r_{\rm g}$, 
and $3.45 r_{\rm g}$,
along the field lines at
$\theta=15^\circ$, $30^\circ$, $45^\circ$, and $60^\circ$, respectively.
Thus, the separation surface is approximately located 
at $5 r_{\rm g} < r < 8 r_{\rm g}$ at $\theta < 30^\circ$.

As for the gap position,
figures~\ref{fig:e2d_1e1}, \ref{fig:e2d_th_1e1}, and
\ref{fig:e2d_1e9} show
that $E_\parallel(r,\theta)$ distribution,
and hence the gap longitudinal extent little depends
on $\theta$,
except near the meridional boundary, 
$\theta \sim \theta_{\rm max}=60^\circ$.
Thus, near the magnetic pole (e.g., $\vert\theta\vert<30^\circ$),
the separation surface is generally located outside the gap
(figs~\ref{fig:width_1e1} \& \ref{fig:width_1e9}). 
This conclusion little depends on the poloidal magnetic field structure,
because the gap solution is essentially determined by the
$\rhoGJ$ distribution, which is governed by the radial component
of the magnetic field near the horizon
(eq.~\ref{eq:rhoGJ_2})
and because the separation surface distribution has a weak dependence
on the poloidal field structure as well
(fig.~2 of HP16).

If a plasma fluid begins to flow with $u_{\rm p} \approx 0$ from the 
separation surface, the plasma particles are accelerated 
up to the Lorentz factor of a few by the MHD interaction
\citep[e.g., eq.~5.2 of][]{hiro92}. 
On the other hand, the electron-positron pairs that cascaded from 
outward-propagating PeV $\gamma$-rays have typical Lorentz factors 
$\sim 10^6$ (for stellar-mass BHs). 
Thus, such pairs continue their outward motion across the separation 
surface, climbing up the \lq hill' of the MHD effective potential $k_0$ 
(fig.~2 of HP16)
very easily. 
In another word, the large charge-to-mass ratio of the electron 
makes the electrostatic acceleration dominate the MHD acceleration, 
which is comparable to the centrifugal or gravitational one. 
Only in the direct vicinity of the horizon, the plasma mass and 
the causality at the horizon makes the gravitational interaction 
overcome the electromagnetic one.
In another word, we can forget about the separation surface
when we consider the formation of a gap in a BH magnetosphere.

It is worth noting that the above-mentioned MHD separation surface
is distinct from what is argued in recent MHD simulations.
In numerical simulations, the flow density decreases to zero
at a specific surface, which requires an imposition of 
\lq density floors'.
Without the density floors, a vacuum would develop in the funnel.
However, such a limitation of numerical MHD 
has nothing to do with the gap formation discussed in the present paper.

\subsection{The case of very small accretion rate}
\label{sec:disc_vacuum}
Let us discuss what is expected when $\dot{m}$ becomes
even less than $\dot{m}_{\rm low}(M)$.
The ADAF photon field peaks around eV for stellar-mass BHs
and around meV for super-massive BHs.
These photons are emitted from the inner-most region,
$r \sim R_{\rm min} \sim 6 r_{\rm g}$,
and decreases outward approximately by $r^{-2}$ law.
Thus, when the gap outer boundary is located at 
$r \gg R_{\rm min}$,
as in the case of $\dot{m} \sim \dot{m}_{\rm low}(R)$,
stationary pair production can be sustained only marginally
in an extended gap.
However, at $\dot{m} < \dot{m}_{\rm low}(R)$,
a stationary pair production can be no longer sustained
and a vacuum region develops in the entire polar funnel.
In this vacuum region, migratory leptons
are accelerated by the vacuum $E_\parallel$ and cascade 
into copious primary electrons and positrons 
accelerated in opposite directions.
Emissions from such cascading primaries will be time-dependent
and may consist of many \lq shots',
whose power spectrum density (PSD) may show some 
characteristic  power law.
In this case, the lower cutoff frequency of the power-law PSD
will be given by the reciprocal of the light crossing time
of the extended gap, which is much longer than the horizon-light-crossing
time scale.

\subsection{Distinction between gap and jet emissions}
\label{sec:disc_distinction}
We finally discuss how to discriminate the gap and jet emissions.
It follows from figure~\ref{fig:SEDa_1e1} that
the gap HE and VHE fluxes increases with decreasing $\dot{m}$.
That is, we can predict an {\it anti-correlation} between the
IR/optical and HE/VHE fluxes.
It forms a contrast to the standard shock-in-jet scenario,
in which the IR/optical and the HE/VHE fluxes will correlate.
Therefore, we propose to simultaneously observe nearby
black hole transients during quiescence
both in near-IR/optical and VHE.
If their time-varying multi-wavelength spectra show 
anti-correlation, it strongly suggests that the photons
are emitted from the BH gap.
For nearby low-luminosity AGNs, the anti-correlation will appear
between submillimeter wavelength and VHE.

In X-rays, the gap emission is very weak.
Thus, if X-ray photons are detected,
they are probably emitted from the jet or from the accretion flow.

Stellar-mass BHs exhibit the strongest gap emission in HE.
Figure~\ref{fig:SEDa_1e1} shows that nearby BH transients
are capable of emitting an order of magnitude greater fluxes
than the LAT detection limit (with 10 years observation).
It means that a HE \lq flare' could be detected by {\it LAT} 
if the flare lasts for a month or so, 
particularly for the sources located away from the galactic plane.
We will examine the plausible sources in a subsequent paper.



\acknowledgments
One of the authors (K. H.) is indebted to 
Dr. T.~Y. Saito for valuable discussion on the CTA sensitivity,
and to Drs. K. Kashiyama, K. Asada, M. Nakamura,
A.~K. Harding, D. Kazanas, S. Shibata
for fruitful discussion. 
This work is supported by the Theoretical Institute for Advanced Research 
in Astrophysics (TIARA) operated under Academia Sinica. 

\end{document}